\documentclass[3p,12pt]{elsarticle}

\usepackage{amssymb}
\usepackage{enumerate}
\usepackage{amsfonts}
\usepackage{amsmath}
\usepackage{amsthm}
\usepackage{psfrag} 
\usepackage{epsfig}
\usepackage{subfigure}
\usepackage{graphicx}
\usepackage{color}
\usepackage{amssymb}
\usepackage{epstopdf}
\usepackage{amsbsy}
\usepackage{latexsym}
\usepackage{moreverb}                      
\usepackage{textcomp}

\usepackage{algorithm}
\usepackage{algorithmic}

\newcommand{\bv}[1]{{\mbox{\boldmath $ #1$}}}
\def\x{{\bv x}}

\newcommand{\lbeq}[1]{{\label{OR:eq:#1}}}
\newcommand{\be}[1]{\begin{equation} \lbeq{#1}}
\newcommand{\ee}{\end{equation}}
\newcommand{\beno}{\begin{equation*}}
\newcommand{\eeno}{\end{equation*}}

\newcommand{\vecbold}[1]{\boldsymbol{#1}}
\renewcommand{\vec}[1]{\vecbold{#1}}

\newtheorem{thm}{Theorem}

\newdefinition{rmk}{Remark}
\newproof{pf}{Proof}
\newproof{pot1}{Proof of Theorem \ref{TM_1}}
\newproof{pot2}{Proof of Theorem \ref{TM_2}}
\newproof{pot3}{Proof of Theorem \ref{TM_3}}
\newproof{pot4}{Proof of Theorem \ref{TM_4}}

\begin{document}

\begin{frontmatter}

\title{Fast Bayesian Optimal Experimental Design\\ for Seismic Source Inversion}

\address[KAUST]{SRI Center for Uncertainty Quantification, King Abdullah University of Science and Technology, Jeddah, Saudi Arabia}
\address[UNM]{Department of Mathematics and Statistics, The University of New Mexico, USA}
\address[UT]{Institute for Computational Engineering and Sciences, The University of Texas at Austin, USA}

\author[KAUST,UT]{Quan Long}
\ead{quan.long@kaust.edu.sa, quan@ices.utexas.edu}

\author[UNM]{Mohammad Motamed}
\ead{motamed@math.unm.edu}

\author[KAUST]{Ra{\'u}l Tempone}
\ead{raul.tempone@kaust.edu.sa}




\begin{abstract}
We develop a fast method for optimally designing experiments in the context of statistical 
seismic source inversion. In particular, we efficiently compute the optimal number and 
locations of the receivers or seismographs. The seismic source is modeled by a point 
moment tensor multiplied by a time-dependent function. The parameters include 
the source location, moment tensor components, and start time and frequency 
in the time function. The forward problem is modeled by elastodynamic wave equations. 
We show that 
the Hessian of the cost functional, which is usually defined as 
the square of the weighted $L_2$ norm of the difference between the experimental 
data and the simulated data, is proportional to the measurement 
time and the number of receivers. Consequently, the posterior distribution of the parameters, 
in a Bayesian setting, concentrates around the ``true'' parameters, 
and we can employ Laplace approximation and speed up the estimation 
of the expected Kullback-Leibler divergence (expected information gain), 
the optimality criterion in the experimental design procedure. 
Since the source parameters span several magnitudes, 
we use a scaling matrix for efficient control of the condition number 
of the original Hessian matrix. We use a second-order accurate finite 
difference method to compute the Hessian matrix 
and either sparse quadrature or Monte Carlo sampling to carry out numerical 
integration. We demonstrate the efficiency, accuracy, and 
applicability of our method on a two-dimensional seismic source inversion problem.
\vskip .1cm
\end{abstract}

\begin{keyword}
Bayesian experimental design \sep Information gain \sep Laplace approximation \sep Monte Carlo sampling \sep Seismic source inversion \sep Sparse quadrature \sep  Uncertainty quantification
\end{keyword}

\end{frontmatter}

\section{Introduction}


In seismic source inversion, the source parameters can be estimated based 
on minimizing a cost functional, which is usually given by the 
weighted $L_2$ norm of the difference between the 
recorded and simulated data. The simulated data are obtained 
by solving a complex forward model, which is described by 
a set of elastic wave equations. The recorded data are usually 
the time series of ground displacements, velocities and accelerations, 
recorded by an array of receivers (seismographs) on the surface of the 
ground and in observation wells. On the other hand, if we treat the 
source parameters as random variables, we seek a complete statistical 
description of all parameter values that are consistent with the noisy measured data. 
This can be achieved using a Bayesian approach \cite{Stuart:10} 
by formulating the inverse problem as a statistical inference problem, 
incorporating uncertainties in the measurements, the forward model, 
and any prior information about the parameters. The solution of this 
inverse problem is the set of posterior probability densities of the 
parameters updated from prior probability densities using Bayes theorem. Meanwhile, the maximum 
a posteriori (MAP) estimation is 
obtained by minimizing a cost functional, 
defined as the negative logarithm of the posterior.

Considering the financial and logistic costs of collecting real data, 
it is important to design an optimal data acquisition procedure, 
with the optimal number and locations of receivers. In the current work, 
we assume that there is additive Gaussian measurement noise and model 
the seismic source by a point moment tensor multiplied by a time-dependent 
function. The parameters include the source location, moment tensor 
components, and start time and frequency in the time function. 
There are in total $N_{\theta}=7$ parameters in the two-dimensional 
model and $N_{\theta}=11$ parameters in a three-dimensional model. 
We then consider the problem of optimal experimental design in a 
Bayesian framework. Under this Bayesian setting, 
a prior probability density function (pdf) of the source parameters 
is given based on expert opinion and/or historical data, and 
the effect of the measured data is incorporated in a likelihood 
function. A posterior pdf of the parameters is then obtained through Bayes theorem by the scaled product 
of the prior pdf and the likelihood function. 
To measure the amount of information obtained from a proposed 
experiment, we use the expected Kullback-Leibler divergence, 
also called the expected information gain. It is specifically defined as the marginalization of the 
logarithmic ratio between the posterior pdf and prior pdf over all possible values of 
seismic source parameters and the data. The optimal experimental setup will then be the
one that maximizes the expected information gain. Finding such an optimal experiment 
requires calculating the expected information gains corresponding to many possible setups.
See \cite{Chaloner:1995} for more details. 

The common method for estimating the expected information gain is based 
on sample averages, which leads to a double-loop integral estimator 
\cite{Huan2013288, Ryan:2003}. This approach can be prohibitively 
expensive when the simulated data are related to solutions
of complex partial differential equations (PDEs). Hence, in such 
cases, such as seismic source inversion, more efficient approaches are required. 

In this paper, we develop a new technique for efficiently computing
the expected information gain of non-repeatable experiments arising 
from seismic source inversion. The efficiency of the new approach, 
which is based on our recent work in \cite{Long:2013}, results from 
the reduction of the double-loop integration to a single-loop one. 
This reduction can be accurately performed by Laplace approximation 
when the posterior distribution of the source parameters is concentrated. 
As the main contribution of the current work, we show that the posterior 
pdf concentrates around ``true'' source parameters, due to the fact that 
the Hessian of the cost functional is proportional to the number of 
receivers and measurement time. Consequently, the error of our approximation 
diminishes by increasing the number of receivers and recording time and by 
improving the precision of our measurements. Hence, we extend the methodology in 
\cite{Long:2013} from repeatable static problems to non-repeatable time-dependent problems of natural earthquakes. From a mathematical point of view, we seek the concentration of posterior probability distribution conditioned on a time series of data instead of repetitive experiments. We also carry out a 
rescaling of the original parameters to address the issue of an ill-conditioned 
Hessian matrix stemming from the large 
span of the parametric magnitudes in the seismic source term. The integrand 
of the approximated expected information gain is a function of the rescaled 
posterior covariance matrix, which can be obtained by solving $N_{\theta}+2$ 
forward problems, consisting of $N_{\theta}+1$ primal problems and 1 dual problem.


The remainder of this paper is organized in the following way. 
In Section 2, we formulate the experimental design problem for seismic source 
inversion and briefly introduce the cost functional and the expected information 
gain for a given experimental setup in the Bayesian setting. 
We present the approximated form of the expected information gain based on Laplace 
approximation and derive the rate of errors in Section 3. In the same section, we also 
summarize the finite difference method for solving the forward problems, the adjoint 
approach for obtaining the Hessian matrix, and the sparse quadratures and Monte Carlo 
sampling for numerical integration. 
In Section 4, we consider numerical examples for a simplified earthquake and design 
optimal experiments. Conclusions are presented in Section 5.


\section{Experimental Design for Seismic Source Inversion}

In this section, we first state the seismic source inversion problem. We then define the expected 
information gain in a Bayesian experimental design framework for seismic inversion.

\subsection{Deterministic full waveform seismic source inversion}

We first consider the full waveform seismic source inversion problem in a 
deterministic setting, which is stated as a PDE-constrained optimization 
problem. The PDEs are given by elastodynamic wave equations in a compact 
spatial domain, $D \subset \mathbb{R}^d$, $d=2, 3$, with a smooth boundary, 
$\partial D$. We augment the PDE with homogeneous initial conditions and 
different types of boundary conditions. The initial-boundary value problem (IBVP) reads: 
\begin{subequations}\label{elastic}
\begin{align}
&\nu(\bold{x}) \, {\bf u}_{tt} (t,\bold{x})- \nabla \cdot \boldsymbol \sigma ({\bf u}(t,\bold{x})) = {\bf f}(t,\bold{x};\boldsymbol{\theta}) \hskip 1cm  \text{in  } [0,T] \times D, \label{elastic_1st}\\
&{\bf u}(0,\bold{x})={\bf 0}, \hskip 6mm {\bf u}_t(0,\bold{x})={\bf 0} \hskip 3.37cm \text{on } \{ t=0 \} \times D, \label{elastic_2nd}\\
&\boldsymbol \sigma({\bf u}(t,\bold{x})) \cdot \hat{\bf n} = {\bf 0} \hskip 5.16cm \text{on } [0,T] \times \partial D_0, \label{elastic_3rd}\\
&{\bf u}_t (t,\bold{x}) = \vec B({\bf x}) \, \boldsymbol \sigma({\bf u}(t,\bold{x})) \cdot \hat{\bf n} \hskip 3.12cm \text{on } [0,T] \times \partial D_1. \label{elastic_4th}
\end{align}
\end{subequations}
The IBVP solution, ${\bf u}= (u_1, \dotsc, u_d)^{\top}$, 
is the displacement field, with $t$ and 
${\bf x}= (x_1, \dotsc, x_d)^{\top}$ the time and location, respectively, 
and $\boldsymbol \sigma$ the stress tensor, which in the case of elastic isotropic materials reads
\begin{equation}\label{stress}
 \boldsymbol \sigma({\bf u}) = \lambda (\bold{x}) \, \nabla \cdot {\bf u} \, I+ \mu (\bold{x}) \, 
 (\nabla {\bf u} + (\nabla {\bf u})^{\top}),
\end{equation}
where $I$ is the identity matrix. The material properties are characterized by 
the density, $\nu$, and the Lam{\'e} parameters, $\lambda$ and $\mu$. Two types 
of boundary conditions are imposed on the boundary, $\partial D$: a homogeneous 
Neumann (stress-free) boundary condition \eqref{elastic_3rd} on $\partial D_0$, 
and an absorbing boundary condition \eqref{elastic_4th} on 
$\partial D_1 = \partial D\setminus \partial D_0$, where ${\hat {\bf n}}$ is 
the outward unit normal to the boundary, and $\vec B$ is a given matrix, see 
for instance \cite{Clayton_Engquist:1977,Petersson_Sjogreen:09}. We visualize such a compact domain together with stress-free and absorbing boundary conditions in Figure \ref{numerics_domain}. The system 
\eqref{elastic} admits longitudinal (P or pressure) and transverse (S or shear) 
waves, which, in the case of constant density, propagate at phase velocities 
$$
c_p = \sqrt{(2 \, \mu+\lambda) / \nu}, \qquad c_s = \sqrt{\mu / \nu},
$$
respectively. There can also be surface waves traveling along a free surface, 
as well as waves that travel along internal material discontinuities.

The function ${\bf f}$ represents the seismic source. We consider the case of a point moment tensor source,
\begin{equation}\label{source}
{\bf f}(t,{\bf x};\boldsymbol{\theta}) = S(t) \, {\bf M} \, \nabla \delta({\bf x} -{\bf x}_s),
\end{equation}
located at ${\bf x}_s = (x_{1s}, \dotsc, x_{ds})^{\top} \in D$, 
where $\nabla \delta$ is the gradient of the Dirac distribution. 
The source time function, $S(t) = S(t;t_s,\omega_s)$, depends on two 
parameters: a time shift, $t_s$, and a frequency parameter, $\omega_s$. 
The moment tensor, ${\bf M}$, is a constant symmetric matrix,
\begin{eqnarray*}
{\bf M}=  \left( \begin{array}{c c c}
m_{x_1  x_1} & \dotsc & m_{x_1  x_d}\\
\vdots & \ddots & \vdots\\
m_{x_1  x_d} & \dotsc & m_{x_d  x_d}
\end{array} \right) \in {\mathbb R}^{d \times d}.
\end{eqnarray*}
The source parameter vector, $\boldsymbol{\theta} \in {\mathbb R}^{N_{\theta}}$, consists of $N_{\theta} = \text{dim}(\boldsymbol{\theta}) = \frac{1}{2} \, d^2+ \frac{3}{2} \, d +2$ parameters,
$$
\boldsymbol{\theta} = (x_{1s}, \dotsc, x_{ds}, t_s, \omega_s, m_{x_1 x_1}, \dotsc, m_{x_d x_d})^{\top}.
$$
There are $N_{\theta}=7$ and $N_{\theta}=11$ parameters, when $d=2$ and $d=3$, respectively.

In the seismic source inversion problem, given $N_R \ge 2$ recorded waveforms, 
the goal is to find the source parameter vector, $\boldsymbol{\theta}$. This is achieved, 
for instance, by minimizing a full waveform cost functional, which is given by the difference 
between the recorded and simulated waveforms,
\begin{equation}\label{cost}
{\mathcal X}(\boldsymbol{\theta}) = \frac{1}{2} \, \sum_{r=1}^{N_R} \int_{t=0}^{T} | {\bf u}(t,{\bf x}_r) - {\bf d}_r(t) |^2 \, dt.
\end{equation}
Here, ${\bf u}(t,{\bf x}_r)$ and ${\bf d}_r(t)$, with $r=1, \dotsc, N_R$, are 
the simulated and recorded waveforms at the $r$-th recording station, respectively, 
and $| {\bf v} |$ denotes the magnitude of the vector, ${\bf v} \in {\mathbb R}^d$. 
In practice, the data are recorded at $N_t$ discrete time levels $t_m \in [0,T]$, 
where $0 = t_0 < t_1 < \dotsc <t_{N_t-2} < t_{N_t-1}=T$. Moreover, 
the problem \eqref{elastic} cannot be solved analytically and needs to 
be discretized. The discrete cost functional corresponding to \eqref{cost} may therefore be written as
\begin{equation}\label{cost_discrete}
{\mathcal X}_d(\boldsymbol{\theta}) = \frac{1}{2} \, \sum_{r=1}^{N_R} \sum_{m=0}^{N_t-1} | {\bf u}_{{\bf i}_r}^m - {\bf d}_r(t_m) |^2,
\end{equation}
where ${\bf u}_{{\bf i}_r}^m \approx {\bf u}(t_m,{\bf x}_r)$ is a finite difference 
approximate solution to \eqref{elastic}, assuming that all recording stations 
coincide with grid points, i.e., ${\bf x}_r = {\bf x}_{{\bf i}_r}$ for some 
$d$-dimensional index vector, ${\bf i}_r = (i_{1r},\dotsc,i_{dr})$. 
See Section \ref{fdm} for more details on the finite difference approximation of \eqref{elastic}.

It is to be noted that the inverse problem based on the minimization 
of \eqref{cost_discrete} is generally an ill-posed problem when the dimension of the parameter vector $N_{\theta}$ is large. This means that infinitely 
many parameter vectors $\boldsymbol{\theta}$ match the recorded data. In such cases, in order to obtain a well-posed problem, the cost functional \eqref{cost_discrete} is often augmented by additional regularizing 
terms. See for instance \cite{Virieux_Operto} for more details. Here, we focus on the case where $N_{\theta}$ is small compared to the number of measurements and the problem is well-posed. 

The above inversion techniques are deterministic 
and do not take into account the uncertainty in the measurements. 
Therefore, in the next section, we consider a Bayesian framework that
accounts for the uncertainty in the problem.

\subsection{Bayesian inference and experimental design for seismic source inversion}

Here, we consider the inversion and the experimental design problems in a Bayesian 
framework by including uncertainty in the form of additive noise in the measurements.    

Let $\boldsymbol{\xi} \in {\mathbb R}^{N_e}$ be a given experimental setup, which 
is a vector of $N_e$ design parameters. For instance, it may consist of the 
number, $N_R$, and the location, $\{ {\bf x}_r \}_{r=1}^{N_R}$, of the seismographs, 
recording the wave forms, $\{{\bf d}_r(t)\}_{r=1}^{N_R}$. Moreover, let 
\begin{equation}\label{forward_model_general}
{\vec g}_r = {\vec g}_r(t,\boldsymbol{\theta},\boldsymbol{\xi}): [0,T] \times 
{\mathbb R}^{N_{\theta}} \times {\mathbb R}^{N_e} \rightarrow {\mathbb R}^d
\end{equation}
be a forward model for computing the vector of outputs at the $r$-th recording station, 
given a source parameter vector, $\boldsymbol{\theta}$, and an experimental 
setup, $\boldsymbol{\xi}$. We further assume that there is additive Gaussian measurement noise 
and collect $N_R$ observation vectors,
\begin{equation} \label{noisy_model}
\vec y_{r}(t)= \vec g_r(t, \vec{\theta}^{*},  \vec\xi) + \vec{\epsilon}_r\,, 
\quad \text{with} \quad r=1,...,N_R,
\end{equation}
using the same experimental set-up, $\vec \xi$. Here, $\vec \theta^{\star}$ 
is the $N_{\theta}$-dimensional vector of "true'' parameters used to 
generate $N_R$ synthetic data, and $\vec\epsilon_r$ is assumed 
to be additive independent and identically distributed (i.i.d.) 
Gaussian noise, $\vec \epsilon_r \sim \mathcal{N}(\vec 0, \vec {\bf C}_{\epsilon})$,
corresponding to the $r$-th observation. Hence, the deterministic output vector,
$\vec g_r( t,\vec \theta^*, \vec\xi)$, is the mean of the observation vector, $\vec y_{r}(t)$. 
Moreover, the collection of $N_R$ observed data points, $\{ \vec y_{r}(t) \}_{r=1}^{N_R}$,
is i.i.d., given specific values of $t$, $\vec \theta^*$, and $\vec\xi$. Note that, 
in practice, the observations are recorded at $N_t$ discrete time levels, $\{ t_m \}_{m=0}^{N_t-1}$.

In seismic inversion, the outputs are usually the waveforms at 
the recording stations obtained by discretizing the forward 
problem \eqref{elastic}. We therefore consider the following 
parameter-to-observable map for the forward model \eqref{forward_model_general}:
\begin{equation}\label{forward_model}
{\bf g}_r = {\bf u}(t_m,{\bf x}_r;\vec \theta, \vec\xi), \quad \text{with} \quad r=1,\dotsc,N_R, \quad m=0,\dotsc,N_t-1,
\end{equation}
where, we have abused the notation to emphasize the 
indirect dependence of ${\bf u}(t_m,{\bf x}_r)$ on $\vec \theta$ and $\vec\xi$. 

The Bayes theorem gives the posterior pdf of the parameter vector as
\begin{align}
 p_{{\cal{\vec\Theta}}}(\vec \theta | \{ \vec y_{r} \}, \vec \xi)  = \frac{p (\{ \vec y_{r} \} | \vec \theta , \vec\xi) \, p_{{\cal{\vec\Theta}}}(\vec \theta)}{p(\{ \vec y_{r} \}| \vec\xi )}.
 \nonumber
\end{align}
We note the fact that $p(\{ \vec y_{r} \}|\vec\xi )$ is a scaling factor, which does not depend on $\vec\theta$. 
The posterior of the parameter vector is proportional to the product of the likelihood function and the prior pdf,
\begin{equation} \label{posterior}
p_{{\cal{\vec\Theta}}}(\vec \theta | \{ \vec y_{r} \}|\vec \xi) \propto \exp \Bigl( - \frac1{2} \, \sum_{r=1}^{N_R} \, 
\sum_{m=0}^{N_t-1}  {\bf r}_r(t_m,\vec \theta, \vec\xi)^{\top} \, {\vec C}_{\epsilon}^{-1} \, {\bf r}_r(t_m,\vec \theta, \vec\xi) 
\Bigr) \, p_{{\cal{\vec\Theta}}}(\vec \theta).
\end{equation}
Here, the residual ${\bf r}_r$ for the $r$-th measurement at the $m$-th time level reads
\begin{equation}\label{res}
{\bf r}_r (t_m,\vec \theta, \vec\xi) :=  \vec y_{r}(t_m, \vec\xi) - \vec g_r(t_m, \vec{\theta}, \vec\xi) = \vec g_r(t_m, \vec{\theta}^{*}, \vec\xi)  - \vec g_r(t_m, \vec{\theta}, \vec\xi)  + \vec{\epsilon}_r\,, 
\end{equation}
where $\vec \theta$ is the generic unknown parameter to evaluate the posterior.

We then define a cost functional as the negative logarithm of the posterior, 
\begin{equation} \label{cost_functional}
{\mathcal L}(\vec{\theta})  := - \log(p_{{\cal{\vec\Theta}}}(\vec \theta | \{ \vec y_{r} \}, \vec\xi)) = 
\frac{1}{2} \sum_{r=1}^{N_R} \, \sum_{m=0}^{N_t-1}   {\bf r}_r(t_m,\vec \theta, \vec\xi)^{\top} \, {\vec C}_{\epsilon}^{-1} \, {\bf r}_r(t_m,\vec \theta, \vec\xi)  - h(\vec \theta) + C,
\end{equation}
where $h(\vec \theta) = \log(p_{{\cal{\vec\Theta}}}(\vec \theta))$, and $C$ is a constant. 

Maximizing the likelihood amounts to minimizing the cost functional 
\eqref{cost_functional}. We also note the direct relation between the 
cost functional \eqref{cost_functional} and the cost functional 
\eqref{cost_discrete} in the deterministic setting, where only the first term in \eqref{cost_functional} 
is retained and ${\vec C}_{\epsilon}$ is an identity matrix. 
The second term in \eqref{cost_functional} is related to the regularizing terms,
which can be added to \eqref{cost_discrete} to obtain a well-posed deterministic problem.
Moreover, we derive our approximation of the expected information gain for a fixed value of $\vec \xi$, which is not treated as a variable in the rest of the paper. We also do not write it as a condition in a probability distribution for the sake of conciseness.

\subsection{Expected information gain}

The Kullback-Leibler divergence is a non-symmetrical measure of the distance between two 
probability distributions \cite{Kullback:1951}. It is related to many 
other statistical invariants. For example, 
if we treat the posterior pdf as a small perturbation of the prior pdf, 
the Hessian of the Kullback-Leibler divergence is the Fisher information matrix. 
The Kullback-Leibler divergence measures how much information the data carries 
about the parameter. When designing an experiment in the context of parameter 
inference, it is important to maximize the expected information gain.

The Kullback-Leibler (K-L) divergence for a given experiment reads
\begin{equation}\label{Dkl}
D_{\text{KL}}( \vec y_r) := \int_{\vec \Theta} \log\frac{p_{{\cal{\vec\Theta}}} (\vec \theta | \{ \vec y_{r}\})}
 {p_{{\cal{\vec\Theta}}} (\vec \theta )} \, p_{{\cal{\vec\Theta}}} (\vec \theta |\{ \vec y_{r}\}) \, d\vec \theta,
\end{equation}
where $p_{{\cal{\vec\Theta}}}(\vec \theta )$ and 
$p_{{\cal{\vec\Theta}}}(\vec \theta | \{ \vec y_r\})$ are respectively the prior and posterior pdfs of the 
unknown random parameter, $\vec \theta$. 

The corresponding expected K-L divergence, which represents the expected 
information gain in the unknown parameters, $\vec \theta$, is then given by:
 \begin{equation}\label{expected_I}
 I :=  {\mathbb E}_{\cal{Y}} [D_{\text{KL}}] = 
 \int_{\mathcal Y} \int_{\vec \Theta} \log\frac{p_{{\cal{\vec\Theta}}} (\vec \theta | \{ \vec y_r\})}
 {p_{{\cal{\vec\Theta}}} (\vec \theta )} \, p_{{\cal{\vec\Theta}}} (\vec \theta |\{ \vec y_r\}) \, p(\{ \vec y_r\})  \, d\vec \theta \,  d\{ \vec y_r\}. 
 \end{equation}
As mentioned in the introduction, using the direct sample average to estimate the expected information gain 
leads to a double-loop summation. In the next section, we present a fast technique 
for computing \eqref{expected_I} based on Laplace approximations, 
motivated by \cite{Long:2013, Long:2014}.

\section{Fast Estimation of Expected Information Gain}
In \cite{Long:2013}, it is shown that if we are able to carry 
out $M$ repetitive experiments and if $M$ is large, the expected 
information gain can be estimated by Laplace approximation 
with a diminishing error asymptotically proportional to $M^{-1}$. 
In seismic source inversion, the large-$M$ assumption is not fulfilled. 
Without losing generality, we show in this section that the error of the 
Laplace approximation also decreases when the number of receivers and 
the measurement time increase. Therefore, we can obtain a fast estimator 
of the expected information gain in the case of non-repeatable experiments.

We consider the cost functional, ${\mathcal L}(\vec{\theta})$,
in \eqref{cost_functional} and let $\hat{\vec \theta}$ be its minimizer,
\begin{equation}\label{arg_min_1}
\hat{\vec \theta} = \text{arg} \, \underset{\vec \theta}{\text{min}} \, {\mathcal L}(\vec{\theta}).
\end{equation}

We further make the following precise assumptions:
\begin{itemize}
\item {\it Assumption A1}. The smallest singular value of the Jacobian of the output model, $\vec g_r(t_m, \vec{\theta})$, 
with respect to $\vec \theta$ is bounded, uniformly in $\vec\theta$ and $t$, from below and away from zero by a constant.

\item {\it Assumption A2}. The output model,
$\vec g_r(t_m, \vec{\theta})$, satisfies $\vec g_r \in {\vec C}^2({\mathbb R}^{N_{\theta}})$, $\forall \, t_m \in {\mathbb R}_+$. 
\end{itemize}

The above two assumptions are used to estimate the magnitudes of quantities, e.g., \eqref{hes_Z_star}, and rates of errors in the approximations, presented in Theorems 1-4 and in the appendix.

We now collect the main results of the paper in Section \ref{thms}, 
followed by the proofs in Section \ref{prfs}. We then present a 
fast numerical method for computing the expected information gain in Section \ref{method}.  

\subsection{Main results}
\label{thms}

In this section, we state the main results. Here, we explain two notations: 
first, for two real vectors ${\bf a}=(a_1,a_2)^{\top}$ and ${\bf b}=(b_1,b_2)^{\top}$, 
we have $\nabla_{\vec\theta} \nabla_{\vec\theta} {\bf a} \circ {\bf b} = \sum_{i=1}^{2}b_i \, 
\nabla_{\vec\theta} \nabla_{\vec\theta} a_i$. Second, ${\mathcal O_P}$ denotes big-O in 
probability, e.g., we write $\varepsilon_N = {\cal{O}}_P\left( N^{-1/2}\right)$, if and only if for a given $\epsilon > 0$, there exist a
constant $N_0$ and an integer $K$, such that for all $N > N_0$, $P(|{\varepsilon}_N| > K N^{-1/2} ) < \epsilon$.

\begin{thm}\label{TM_1}
Under assumptions A1 and A2, the minimizer \eqref{arg_min_1} of the cost functional \eqref{cost_functional} is given by
\begin{equation}\label{1st_aprox}
\hat{\vec \theta} = {\vec \theta}^* + {\mathcal O}_P({N}^{-1/2}),
\end{equation}
or by
\begin{equation}\label{2nd_aprox} 
\hat{\vec \theta} = {\vec \theta}^* + 
{{\vec H}({\vec \theta}^*)}^{-1} \, \Bigl( \sum_{r=1}^{N_R} \, \sum_{m=0}^{N_t-1} \,  {\vec{\epsilon}_r}^{\top} \, 
{\vec C}_{\epsilon}^{-\top}  \, \nabla_{\vec\theta} \vec g_r(t_m, {\vec{\theta}}^*) \,  
+ \nabla_{\vec\theta} h({\vec\theta}^*) \Bigr)^{\top}+ {\mathcal O}_P({N}^{-1}),
\end{equation}
where the total number of measurements is 
$$N = N_R\times N_t\, ,$$
and the Hessian of $\vec g_r$ with respect to $\vec\theta$ is
$$
{\vec H} ({\vec \theta}^*)=
 \sum_{r=1}^{N_R} \, \sum_{m=0}^{N_t-1}  \, \Bigl( \nabla_{\vec\theta} \vec g_r(t_m, {\vec{\theta}}^*)^{\top} \, {\vec C}_{\epsilon}^{-1} \,  \nabla_{\vec\theta} \vec g_r(t_m, {\vec{\theta}}^*)- 
\,  \nabla_{\vec\theta} \nabla_{\vec\theta} \vec g_r(t_m, {\vec{\theta}}^*) \circ {\vec C}_{\epsilon}^{-1} \, 
\vec{\epsilon}_r   \Bigr) - \nabla_{\vec\theta} \nabla_{\vec\theta} h({\vec \theta}^*).
$$
\end{thm}

\begin{thm}\label{TM_2}
(Gaussian approximation of the posterior) The posterior pdf in \eqref{posterior} 
can be approximated by a Gaussian pdf as follows:
\begin{align} \label{Gauss_posterior}
p_{{\cal{\vec\Theta}}}(\vec \theta | \{ \vec y_{r} \}) =&\frac{\exp \Bigl( - \frac1{2} \, (\vec \theta - \hat{\vec\theta})^{\top} \,   {\vec H}(\hat{\vec\theta})     \, (\vec \theta - \hat{\vec\theta}) \Bigr)}
{(2 \, \pi)^{N_{\theta}/2} \, | {\vec H}(\hat{\vec\theta}) |^{1/2}} 
\exp \left[\mathcal O_P\left(|\vec\theta -\hat{\vec\theta}|^3\right) \right]\\
=&\tilde{p}_{{\cal{\vec\Theta}}}(\vec \theta | \{ \vec y_r \}) 
+ \mathcal O_P \left(|\vec\theta -\hat{\vec\theta}|^3 \right),
\end{align}
where
\begin{equation}\label{Hessian_Y}
{\vec H}(\hat{\vec\theta})  = {\vec H}_1(\hat{\vec\theta}) + {\vec H}_2(\hat{\vec\theta})
- \nabla_{\vec\theta} \nabla_{\vec\theta} h(\hat{\vec\theta}), 
\end{equation}
with
\begin{align}
{\vec H}_1(\hat{\vec\theta})  &=  \sum_{r=1}^{N_R} \, \sum_{m=0}^{N_t-1}  
\nabla_{\vec\theta} \vec g_r(t_m, \hat{\vec\theta})^{\top} \, {\vec C}_{\epsilon}^{-1}\,    
\nabla_{\vec\theta} \vec g_r(t_m, \hat{\vec\theta}) = {\mathcal O}(N),  \label{H1}\\
{\vec H}_2(\hat{\vec\theta})  &= -  \sum_{r=1}^{N_R} \, \sum_{m=0}^{N_t-1}  \,  
\nabla_{\vec\theta} \nabla_{\vec\theta} \vec g_r(t_m, \hat{\vec\theta}) 
\circ {\vec C}_{\epsilon}^{-1} \, \bigl( {\vec g}_r(t_m,{\vec\theta}^*) -  {\vec g}_r(t_m,\hat{\vec\theta})  \bigr) \nonumber\\   
& \quad \, - \sum_{r=1}^{N_R} \, \sum_{m=0}^{N_t-1}  \,  \nabla_{\vec\theta}
\nabla_{\vec\theta} \vec g_r(t_m, \hat{\vec\theta}) \circ {\vec C}_{\epsilon}^{-1} \,  {\vec\epsilon}_r 
={\mathcal O}_P({N}^{1/2}).\label{H2}
\end{align}
\end{thm}
\begin{thm}\label{TM_3}
Under assumptions A1 and A2, the expected information gain is given by 
\begin{multline}\label{eq:miMultiD1} 
 I = \int_{{\vec{\Theta}}} \int_{{\mathcal{Y}}} \left[  - \frac{1}{2}\log((2\pi)^{N_{\theta}} \, |{{\vec H}(\hat{\vec\theta})}^{-1}|) 
 - \frac{N_{\theta}}{2} - h(\hat{\vec\theta}) - 
\frac{\text{tr}({{\vec H}(\hat{\vec\theta})}^{-1} \, \nabla\nabla h(\hat{\vec\theta}))}{2} 
\right] \\
p(\{\vec y_r\}|\vec\theta^*) \, d\{\vec y_r\} \, p(\vec\theta^*) \, d\vec\theta^* 
+ {\mathcal O}\left({N}^{-2}\right),
\end{multline}
where ${\vec H}(\hat{\vec\theta})$ is given by \eqref{Hessian_Y}-\eqref{H2}.
\end{thm}

\begin{thm}\label{TM_4}
Under assumptions A1 and A2, the expected information gain is given by
\begin{equation}\label{eq:miMultiD2} 
 I = \int_{{\vec{\Theta}}}\hat{D}_{KL}(\vec\theta^*)
 p(\vec\theta^*) \, d\vec\theta^* + {\mathcal O}\left({N}^{-1}\right)\, ,
\end{equation}
with
\begin{equation}\label{DKL_hat}
\hat{D}_{KL} = - \frac{1}{2}\log((2 \pi)^{N_{\theta}}| {{\vec H}_1({\vec\theta}^*)}^{-1}|) 
- \frac{N_{\theta}}{2} - h(\vec\theta^*).
\end{equation}
Here, ${\vec H}_1({\vec\theta}^*)$ is given by \eqref{H1} with $\hat{\vec\theta}$ replaced by ${\vec\theta}^*$, which is the ``true'' 
parameter generating the synthetic data, cf. \eqref{noisy_model}
\end{thm}

\begin{rmk}
The expected information gain, $I$, in \eqref{expected_I} can be 
approximated by the integral term in \eqref{eq:miMultiD1} with an 
asymptotic error proportional to ${N}^{-2}$. The expected information gain
can also be approximated by the integral term in \eqref{eq:miMultiD2} with an
asymptotic error proportional to ${N}^{-1}$. The 
dimension of the integration domain in Theorem \ref{TM_4} is less than that in Theorem \ref{TM_3}. 
\end{rmk}

\subsection{Proof of the main results}
\label{prfs}

In this section, we collect the proofs of the main results.

\begin{pot1}
We first find ${\mathcal Z}(\vec\theta)$ so that 
\begin{equation}\label{Y_Z}
\hat{\vec \theta} = \text{arg} \, \underset{\vec \theta}{\text{min}} \, {\mathcal L}(\vec{\theta}) = \text{arg} \, \underset{\vec \theta}{\text{min}} \, {\mathcal Z}(\vec{\theta}).
\end{equation}
By \eqref{res} and  \eqref{cost_functional}, we have
\begin{equation}\label{LZ}
{\mathcal L}(\vec{\theta}) = {\mathcal Z}(\vec{\theta}) + \frac{1}{2} \sum_{r=1}^{N_R} \, \sum_{m=0}^{N_t-1} \,    \vec{\epsilon}_r^{\top} \, {\vec C}_{\epsilon}^{-1} \,  \vec{\epsilon}_r  + C,
\end{equation}
where
\begin{align*}
{\mathcal Z}(\vec{\theta}) =& 
\frac{1}{2}  \sum_{r=1}^{N_R} \, \sum_{m=0}^{N_t-1}  \,  \bigl( \vec g_r(t_m, \vec{\theta}^{*}) - \vec g_r(t_m, \vec{\theta}) \bigr)^{\top} \, {\vec C}_{\epsilon}^{-1} \,    \bigl( \vec g_r(t_m, \vec{\theta}^{*}) - \vec g_r(t_m, \vec{\theta}) \bigr)  +\\
& \sum_{r=1}^{N_R} \, \sum_{m=0}^{N_t-1}  \,  \bigl( \vec g_r(t_m, \vec{\theta}^{*}) - \vec g_r(t_m, \vec{\theta}) \bigr)^{\top} \, {\vec C}_{\epsilon}^{-1} \,  \vec{\epsilon}_r  - h(\vec \theta). 
\end{align*}
Note that the last two terms in \eqref{LZ} are independent of $\vec\theta$. We then have
\begin{multline}\label{grad_Z}
\nabla_{\vec\theta} {\mathcal Z}(\vec{\theta}) = 
-  \sum_{r=1}^{N_R} \, \sum_{m=0}^{N_t-1}  \,  \bigl( \vec g_r(t_m, \vec{\theta}^{*}) - \vec g_r(t_m, \vec{\theta}) \bigr)^{\top} \, {\vec C}_{\epsilon}^{-1} \, \nabla_{\vec\theta} \vec g_r(t_m, \vec{\theta})  \,   -\\
 \sum_{r=1}^{N_R} \, \sum_{m=0}^{N_t-1} \,  {\vec{\epsilon}_r}^{\top} \,  {\vec C}_{\epsilon}^{-\top}  \, \nabla_{\vec\theta} \vec g_r(t_m, \vec{\theta}) \,   - \nabla_{\vec\theta} h(\vec \theta),
\end{multline}
where 
\begin{align}\label{hes_Z}
\nabla_{\vec\theta} \nabla_{\vec\theta} {\mathcal Z}(\vec{\theta}) =& 
-  \sum_{r=1}^{N_R} \, \sum_{m=0}^{N_t-1} \,  \nabla_{\vec\theta} \nabla_{\vec\theta} \vec g_r(t_m, \vec{\theta}) \circ {\vec C}_{\epsilon}^{-1} \,    \bigl( \vec g_r(t_m, \vec{\theta}^{*}) - \vec g_r(t_m, \vec{\theta}) \bigr) + \nonumber \\ 
& \sum_{r=1}^{N_R} \, \sum_{m=0}^{N_t-1}  \,  \nabla_{\vec\theta} \vec g_r(t_m, \vec{\theta})^{\top} \, {\vec C}_{\epsilon}^{-1} \,  \nabla_{\vec\theta} \vec g_r(t_m, \vec{\theta})- \nonumber \\
&\sum_{r=1}^{N_R} \, \sum_{m=0}^{N_t-1} \,  \nabla_{\vec\theta} \nabla_{\vec\theta} \vec g_r(t_m, \vec{\theta}) \circ {\vec C}_{\epsilon}^{-1} \,  \vec{\epsilon}_r   - \nabla_{\vec\theta} \nabla_{\vec\theta} h(\vec \theta).
\end{align}
The Taylor expansion of $\nabla_{\vec\theta} {\mathcal Z}(\vec{\theta})$ in \eqref{grad_Z} around ${\vec\theta}^*$ reads
\begin{equation}\label{Taylor1}
\nabla_{\vec\theta} {\mathcal Z}(\vec{\theta}) = \nabla_{\vec\theta} {\mathcal Z}({\vec\theta}^*) + {(\vec\theta - {\vec\theta}^*)}^{\top} \, \nabla_{\vec\theta} \nabla_{\vec\theta} {\mathcal Z}({\vec\theta}^*) + {\mathcal O} (|\vec\theta - {\vec\theta}^* |^2).
\end{equation}
We now evaluate \eqref{Taylor1} at 
$\vec\theta = \hat{\vec\theta}$, and noting that $\nabla_{\vec\theta} {\mathcal Z}(\hat{\vec\theta}) = 0$ 
by \eqref{Y_Z}, we write
\begin{equation}\label{diff}
\hat{\vec\theta} - {\vec\theta}^* = - \bigl( \nabla_{\vec\theta} \nabla_{\vec\theta} {\mathcal Z}({\vec\theta}^*)  \bigr)^{-1} \,  {\nabla_{\vec\theta} {\mathcal Z}({\vec\theta}^*)}^{\top} + {\mathcal O} (|\hat{\vec\theta} - {\vec\theta}^* |^2).
\end{equation}
From \eqref{grad_Z}, we obtain
\begin{equation}\label{grad_Z_star}
\nabla_{\vec\theta} {\mathcal Z}({\vec\theta}^*) = 
-  \sum_{r=1}^{N_R} \, \sum_{m=0}^{N_t-1} \,  {\vec{\epsilon}_r}^{\top} 
\,  {\vec C}_{\epsilon}^{-\top}  \, \nabla_{\vec\theta} 
\vec g_r(t_m, {\vec{\theta}}^*) \,   - \nabla_{\vec\theta} h({\vec\theta}^*) = {\mathcal O}_P({N}^{1/2}),
\end{equation}
where the order, ${\mathcal O}_P({N}^{1/2})$, follows Appendix A.  

Moreover, from \eqref{hes_Z}, we obtain
\begin{align}\label{hes_Z_star}
\nabla_{\vec\theta} \nabla_{\vec\theta} {\mathcal Z}({\vec\theta}^*) &=
 \sum_{r=1}^{N_R} \, \sum_{m=0}^{N_t-1}  \,  \nabla_{\vec\theta} \vec g_r(t_m, {\vec{\theta}}^*)^{\top} \, {\vec C}_{\epsilon}^{-1} \,  \nabla_{\vec\theta} \vec g_r(t_m, {\vec{\theta}}^*)- \nonumber \\
&\quad \, \sum_{r=1}^{N_R} \, \sum_{m=0}^{N_t-1} \,  \nabla_{\vec\theta} \nabla_{\vec\theta} \vec g_r(t_m, {\vec{\theta}}^*) \circ {\vec C}_{\epsilon}^{-1} \,  \vec{\epsilon}_r   - \nabla_{\vec\theta} \nabla_{\vec\theta} h({\vec \theta}^*) \nonumber\\
&= {\mathcal O}(N) + {\mathcal O}_P({N}^{1/2}) = {\mathcal O}_P(N),
\end{align}
where the orders ${\mathcal O}_P({N}^{1/2})$ and ${\mathcal O}(N)$ of 
the second and first terms in the right-hand side follow Appendices 
B and C, respectively. The proof is completed by \eqref{diff}-\eqref{hes_Z_star}.   \hfill$\Box$
\end{pot1}

\begin{pot2}
Let $\tilde{\mathcal L}(\vec\theta)$ be the second-order Taylor 
expansion of ${\mathcal L}(\vec\theta)$ around $\hat{\vec\theta}$,
$$
\tilde{\mathcal L}(\vec\theta) = {\mathcal L}(\hat{\vec\theta}) +  \nabla_{\vec\theta} 
{\mathcal L}(\hat{\vec\theta}) \, (\vec \theta - \hat{\vec\theta}) + \frac1{2} \, (\vec \theta - \hat{\vec\theta})^{\top} \,  \nabla_{\vec\theta}  \nabla_{\vec\theta} {\mathcal L}(\hat{\vec\theta}) \, (\vec \theta - \hat{\vec\theta}).
$$
The first term in the right hand side is independent of $\vec\theta$. 
Moreover, by \eqref{Y_Z}, the second term is zero, since $\nabla_{\vec\theta} {\mathcal L}(\hat{\vec\theta}) = 0$. We are 
therefore left only with the third term. 
Similar to the proof of Theorem \ref{TM_1}, \eqref{Hessian_Y} follows easily. 
Furthermore, the growth order in \eqref{H1} follows in a similar way to Appendix C. 
It is left to show \eqref{H2}. By \eqref{res}, and similar to the proof of Theorem \ref{TM_1}, we write
\begin{align}\label{H2_proof}
{\vec H}_2(\hat{\vec\theta}) =& - \sum_{r=1}^{N_R} \, \sum_{m=0}^{N_t-1}  \,  \nabla_{\vec\theta} \nabla_{\vec\theta} \vec g_r(t_m, \hat{\vec\theta}) \circ {\vec C}_{\epsilon}^{-1} \,  {\vec\epsilon}_r  - \nonumber \\
&  \sum_{r=1}^{N_R} \, \sum_{m=0}^{N_t-1}  \,  \nabla_{\vec\theta} \nabla_{\vec\theta} \vec g_r(t_m, \hat{\vec\theta}) \circ {\vec C}_{\epsilon}^{-1} \, \bigl( {\vec g}_r(t_m,{\vec\theta}^*) -  {\vec g}_r(t_m,\hat{\vec\theta})  \bigr).
\end{align}
The first term in the right-hand side of \eqref{H2_proof} is of 
order ${\mathcal O}_P({N}^{1/2})$, similar to Appendix B. For the second 
term in the right-hand side of \eqref{H2_proof}, we use Taylor expansion and write
$$
\vec g_r(t_m,{\vec\theta}) = \vec g_r(t_m,{\vec\theta}^*) + \nabla_{\vec\theta} \vec g_r(t_m,{\vec\theta}^*) \, (\vec\theta - {\vec\theta}^*) + {\mathcal O} (|\vec\theta - {\vec\theta}^* |^2).
$$
Then at ${\vec\theta} = \hat{\vec\theta}$, using \eqref{1st_aprox}, we have
$$
\vec g_r(t_m,\hat{\vec\theta}) - \vec g_r(t_m,{\vec\theta}^*) = {\mathcal O}_P({N}^{-1/2}). 
$$
The first term in the right-hand side of \eqref{H2_proof} dominates, and this completes the proof. \hfill$\Box$
\end{pot2}

\begin{pot3}
We first rewrite the information gain \eqref{Dkl} as
\begin{align*}
 D_{KL} =& \int_{\vec\Theta}\log \left( \frac{p_{{\cal{\vec\Theta}}}(\vec\theta | \{\vec y_r\})}{p_{{\cal{\vec\Theta}}}(\vec\theta)} \right)
 \tilde{p}_{{\cal{\vec\Theta}}}(\vec\theta | \{\vec y_r\}) d\vec\theta +\\
 & \int_{\vec\Theta}\log \left( \frac{p_{{\cal{\vec\Theta}}}(\vec\theta | \{\vec y_r\})}
 {p_{{\cal{\vec\Theta}}}(\vec\theta) } \right)
 ( p_{{\cal{\vec\Theta}}}(\vec\theta | \{\vec y_r\}) - \tilde{p}_{{\cal{\vec\Theta}}}(\vec\theta | \{\vec y_r\}) )d\vec\theta,
\end{align*}
where $\tilde{p}$ is the Gaussian approximation of the posterior, $p$, 
given in Theorem \ref{TM_2} by \eqref{Gauss_posterior}. Then, we can write
\begin{equation}\label{eq:dkl}
D_{KL} = D_{1} + D_{2} + D_{3} + D_{4},
\end{equation}
where
\begin{align}
 &D_{1} := \int_{\vec\Theta}\log \left(\tilde{p}_{{\cal{\vec\Theta}}}(\vec\theta | \{\vec y_r\}) \right)
 \tilde{p}_{{\cal{\vec\Theta}}}(\vec\theta | \{\vec y_r\}) d\vec\theta, \label{D_1}\\
 &D_{2} :=  - \int_{\vec\Theta}\log \left( p_{{\cal{\vec\Theta}}}(\vec\theta) \right)
 \tilde{p}_{{\cal{\vec\Theta}}}(\vec\theta | \{\vec y_r\}) d\vec\theta = 
 - \int_{\vec\Theta} h(\vec\theta) \tilde{p}_{{\cal{\vec\Theta}}}(\vec\theta | \{\vec y_r\}) d\vec\theta, \label{D_2} \\
&D_{3} :=  \int_{\vec\Theta}\log \left( \frac{p_{{\cal{\vec\Theta}}}(\vec\theta | \{\vec y_r\})}
 {\tilde{p}_{{\cal{\vec\Theta}}}(\vec\theta | \{\vec y_r\}) } \right)
 \tilde{p}_{{\cal{\vec\Theta}}}(\vec\theta | \{\vec y_r\}) d\vec\theta,  \label{D_3}\\
 &D_{4} := \int_{\vec\Theta}\log \left( \frac{p_{{\cal{\vec\Theta}}}(\vec\theta | \{\vec y_r\})}
 {p_{{\cal{\vec\Theta}}}(\vec\theta) } \right) ( p_{{\cal{\vec\Theta}}}(\vec\theta | \{\vec y_r\}) 
 - \tilde{p}_{{\cal{\vec\Theta}}}(\vec\theta | \{\vec y_r\}) )d\vec\theta. \label{D_4}
\end{align}

The first term \eqref{D_1} reads
\begin{equation}\label{eq:firstterm}
 D_1 = -\frac{1}{2}\log((2\pi)^{N_{\theta}} | \tilde{\vec C} |) - \frac{N_{\theta}}{2}, \quad \tilde{\vec C} := {\vec H}(\hat{\vec\theta})^{-1}.
\end{equation}

For the second term \eqref{D_2}, we first 
Taylor expand $h(\vec\theta) = \log(p_{{\cal{\vec\Theta}}}(\vec\theta))$ about $\hat{\vec\theta}$ to obtain
$$
 h(\vec\theta) = \sum_{|\vec\alpha|\leq 4} \frac{D^{\vec\alpha} h(\hat{\vec\theta})}{\vec\alpha !} 
 (\vec\theta - \hat{\vec\theta})^{\vec\alpha} 
 + \mathcal O_P( | \vec\theta - \vec{\hat\theta}|^5 ), 
$$
where $\vec\alpha \in {\mathbb N}^{N_{\theta}}$ is a multi-index with the following properties: 
$$
|\vec\alpha|= \sum_{i=1}^{N_{\theta}} \alpha_i, \qquad  
\vec\alpha! = \prod_{i=1}^{N_{\theta}} \alpha_i !, \qquad 
(\vec \theta)^{\vec\alpha} = \prod_{i=1}^{N_{\theta}} \theta_i^{\alpha_i}.
$$
The odd central moments of the multivariate Gaussian, $\tilde{p}$, in \eqref{Gauss_posterior} are zero, and 
the parameter posterior covariance, $\tilde{\vec C}$, is of order ${\mathcal O}_P({N}^{-1})$, 
due to Theorem \ref{TM_2}. Moreover, the fourth central moment 
of this multivariate Gaussian is a quadratic form of the second moment, 
and hence is of order $\mathcal O_P({N}^{-2})$.  Consequently, we have
\begin{align} \label{eq:secondterm}
 D_2 &= - \int_{\vec\Theta} \left[\sum_{|\vec\alpha|\leq 4} \frac{D^{\vec\alpha} h(\hat{\vec\theta})}{\vec\alpha !} 
 (\vec\theta - \hat{\vec\theta})^{\vec\alpha} 
 + \mathcal O_P( | \vec\theta - \vec{\hat\theta}|^5 )\right] \tilde{p}(\vec\theta | \{\vec y_r\}) d\vec\theta \nonumber\\
 &=  -h(\hat{\vec\theta}) - \frac{{\tilde{ \vec C} : 
 \nabla\nabla h(\hat{\vec\theta})}}{2} + \mathcal O_P({N}^{-2}).
\end{align}
%
Here, $\vec A : \vec B = \sum_{i,j}{A}_{ij}{B}_{ij}$ is the component-wise 
inner product of two matrices, $\vec A = (A_{ij})$ and $\vec B = (B_{ij})$ of the same size.

%

Next, we consider the third term \eqref{D_3}. Since the approximate 
posterior, $\tilde{p}$, is a second-order Taylor approximation of the log posterior, $p$, we have

\begin{equation*}
\log \left( \frac{p_{{\cal{\vec\Theta}}}(\vec\theta | \{\vec y_r\})}
 {\tilde{p}_{{\cal{\vec\Theta}}}(\vec\theta | \{\vec y_r\}) } \right) =  \sum_{|\vec\alpha|=3} \frac{D^{\vec\alpha} 
 h_p(\hat{\vec\theta})}{\vec\alpha !} (\vec\theta - 
 \hat{\vec\theta})^{\vec\alpha} + \mathcal O_P(| \vec\theta - \vec{\hat\theta} |^4), 
 \nonumber
\end{equation*}
where $h_p = \log(p_{{\cal{\vec\Theta}}}(\vec\theta | \{\vec y_r\}))$. 
Similar to the analysis of the second term, $D_2$, we can easily show that
\begin{equation}\label{eq:thirdterm}
D_3 =\mathcal O_P({N}^{-2}).
\end{equation}

Finally, for the fourth term \eqref{D_4}, we have 
\begin{multline*}
 D_4 =\int_{\cal{\vec\Theta}} \log \left( \frac{p_{{\cal{\vec\Theta}}}(\vec\theta | \{\vec y_r\})}
 {p_{{\cal{\vec\Theta}}}(\vec\theta) } \right) \\
 \biggl\{ \exp\biggl[\sum_{|\vec\alpha|=3} 
 \frac{D^{\vec\alpha} h_p(\hat{\vec\theta})}{\vec\alpha !} (\vec\theta - \hat{\vec\theta})^{\vec\alpha} 
 + \mathcal O_P(|\vec\theta - \hat{\vec\theta}|^4)\biggr]
 -1\biggr\} 
 \tilde{p}_{{\cal{\vec\Theta}}}(\vec \theta|\{\vec y_r\}) d\vec\theta.
\end{multline*}
After the first-order Taylor expansion of the exponential term, we obtain 
$$
D4=\int_{\cal{\vec\Theta}} \log \left( \frac{p_{{\cal{\vec\Theta}}}(\vec\theta | \{\vec y_r\})}
 {p_{{\cal{\vec\Theta}}}(\vec\theta) } \right) 
 \biggl\{ \sum_{|\vec\alpha|=3} 
 \frac{D^{\vec\alpha} h_p(\hat{\vec\theta})}{\vec\alpha !} (\vec\theta - \hat{\vec\theta})^{\vec\alpha} 
 + \mathcal O_P(|\vec\theta - \hat{\vec\theta}|^4)\biggr\} \, 
 \tilde{p}_{{\cal{\vec\Theta}}}(\vec\theta|\{\vec y_r\}) d\vec\theta.
$$
Since $\log\left( \frac{p_{{\cal{\vec\Theta}}}(\hat{\vec\theta} | \{\vec y_r\})}
 {p_{{\cal{\vec\Theta}}}(\hat{\vec\theta}) } \right)$ is asymptotically $\mathcal O_P\left(\log(N)\right)$,  
 and the third moment of a multivariate Gaussian is zero, we have
 \begin{equation}\label{eq:fourthterm}
D_4 = \int_{\cal{\vec\Theta}}\log \left( \frac{p_{{\cal{\vec\Theta}}}(\vec\theta | \{\vec y_r\})}
 {p_{{\cal{\vec\Theta}}}(\vec\theta) } \right) \mathcal O_P(|\vec\theta - \hat{\vec\theta}|^4) \,
\tilde{p}_{{\cal{\vec\Theta}}}(\vec\theta|\{\vec y_r\}) \, d\vec\theta = \mathcal O_P({N}^{-2}\log(N)).
\end{equation}
This is the fourth moment of the Gaussian posterior, $\tilde{p}$, which has already 
been shown to be inversely proportional to ${N}^2$.

Substituting \eqref{eq:firstterm}-\eqref{eq:fourthterm} into \eqref{eq:dkl}, we obtain
\begin{equation}\label{eq:dkl1}
 D_{KL} = -\frac{1}{2}\log((2\pi)^{N_{\theta}} |\tilde{\vec C}|) - \frac{N_{\theta}}{2} 
 + h(\hat{\vec\theta}) + \frac{\tilde{ \vec C} : \nabla\nabla h(\hat{\vec\theta})}{2}
 + \mathcal O_P\left({N}^{-2}\log(N)\right).
\end{equation}
After marginalization over data, and conditioning on the ``true'' parameter, which has a pdf identical to the prior 
of the unknown parameters, the approximation and error estimation for the expected information gain \eqref{eq:miMultiD1} is obtained. This completes the proof.  \hfill$\Box$
\end{pot3}

\begin{pot4}
Using Woodbury's formula \cite{Hager:1989} to invert \eqref{Hessian_Y}, we have 
\begin{align}
{\vec H}(\hat{\vec\theta})^{-1}  = &
\left[{\vec H}_1(\hat{\vec\theta}) + {\vec H}_2(\hat{\vec\theta})  
- \nabla_{\vec\theta} \nabla_{\vec\theta} h(\hat{\vec\theta})\right]^{-1} \nonumber\\
= & {\vec H}_1(\hat{\vec\theta})^{-1} 
+ {\vec H}_1(\hat{\vec\theta})^{-1}\vec L (\vec\Lambda^{-1} + \vec R{\vec H}_1(\hat{\vec\theta})^{-1} \vec L )^{-1} 
\vec R  {\vec H}_1(\hat{\vec\theta})^{-1} \label{eq:woodbury}
\end{align}
with ${\vec H}_2(\hat{\vec\theta}) - \nabla_{\vec\theta} \nabla_{\vec\theta} h(\hat{\vec\theta}) = 
\vec L \vec\Lambda \vec R$ the corresponding eigenvalue decomposition. The second term on the right-hand side of 
\eqref{eq:woodbury} is of order $\mathcal O_P({N}^{-\frac{3}{2}})$. Hence,

\begin{equation*}
{\vec H}(\hat{\vec\theta})^{-1} = \vec H_1(\hat{\vec\theta})^{-1} + \mathcal O_P({N}^{-\frac{3}{2}}).
\end{equation*}
We therefore can approximate the information gain, $D_{KL}$, in \eqref{eq:dkl1} by 
\begin{equation*}
D_{KL} =  \hat{D}_{KL} + \mathcal O_P({N}^{-\frac{3}{2}}), 
\end{equation*}
where
\begin{equation*}
 \hat{D}_{KL}  := -\frac{1}{2}\log((2\pi)^{N_{\theta}} |\vec H_1(\hat{\vec\theta})^{-1}|) - \frac{N_{\theta}}{2} 
 + h(\hat{\vec\theta}) + \frac{ \vec H_1(\hat{\vec\theta})^{-1}: \nabla\nabla h(\hat{\vec\theta})}{2}\,.
\end{equation*}

Next, we Taylor expand $\hat{D}_{KL}$ about $\vec\theta^*$, 
\begin{align}
 \hat{D}_{KL}(\hat{\vec\theta }) = \hat{D}_{KL}(\vec\theta^*) + \nabla \hat{D}_{KL}(\vec\theta^*)^T
 (\hat{\vec\theta} - \vec\theta^*) + \mathcal O_P(| \hat{\vec\theta} - \vec\theta^*|^2)
\end{align}
Now, by rearranging the expected information gain, we get
\begin{equation}\label{eq:theorem3}
 I = \int_{\vec\Theta} \hat{D}_{KL}(\vec\theta^*)p(\vec \theta^*) d\vec \theta^*  
 +  \int_{\vec\Theta}\int_{\{\mathsf{\vec\epsilon_r}\}} \nabla \hat{D}_{KL}(\vec\theta^*)^T 
 (\hat{\vec\theta} - \vec\theta^*) d\vec \theta^* 
 p(\{\vec\epsilon_r\}) d \{\vec\epsilon_r\}
 +  \mathcal O({N}^{-1}).
\end{equation}
By Theorem \ref{TM_1}, as $N \rightarrow \infty$, the difference, $\hat{\vec\theta} - \vec\theta^*$, 
is asymptotically proportional to $\vec\epsilon_r$ (see \eqref{2nd_aprox}); in other words, 
$\sum_{r=1}^{N_R} \, \sum_{m=0}^{N_t-1} \,  {\vec{\epsilon}_r}^{\top} \, 
{\vec C}_{\epsilon}^{-\top}  \, \nabla_{\vec\theta} \vec g_r(t_m, {\vec{\theta}}^*) = {\mathcal O}_P \left(
N^{1/2}\right)$ is dominant over $\nabla_{\vec\theta} h({\vec\theta}^*) = {\mathcal O} \left(1\right)$, and 
${\vec H}(\hat{\vec\theta})$ is dominated by its deterministic component, ${\vec H}_1(\hat{\vec\theta})$ (see 
Theorem \ref{TM_2}).
Since $\vec\epsilon_r$ is a centered Gaussian random vector, the second integral on the right-hand side of \eqref{eq:theorem3}
vanishes. Therefore, the approximated expected information gain as shown in \eqref{eq:theorem3} has 
an error of order ${N}^{-1}$, and the proof is complete.  \hfill$\Box$
\end{pot4}

\subsection{Fast numerical approach}
\label{method}

\subsubsection{Finite difference approximation of the problem}
\label{fdm}

Consider a rectangular spatial domain, $D = [-L_1,L_1] \times [-L_2,0]$, in ${\mathbb R}^2$. 
We employ a second-order accurate finite difference scheme, proposed in \cite{Nilsson_etal:07},
for solving \eqref{elastic}. Let $h = 2 \, L_1/ (N_1-1) = L_2/ (N_2-1) > 0$ denotes the 
spatial grid-length, where $N_1$ and $N_2$ are the numbers of grid points in the 
$x_1$ and $x_2$ directions, respectively. Let ${\bf i} = (i,j)$, and for 
$i=1,\ldots, N_1$ and $j=1,\ldots, N_2$, consider the computational grid
$$
{\bf x}_{\bf i} = (x_{1i},x_{2j}) =(-L_1+ (i-1) \, h, -L_2+(j-1) \, h).
$$
Denote by ${\bf u}_{\bf i}(t) = (u_{1 {\bf i}}(t),u_{2 {\bf i}}(t))^{\top}$ 
the semi-discrete approximation of ${\bf u}(t,{\bf x}_{\bf i})$. 
For interior grid points and grid points on the free surface boundary 
($\partial D_0 = \{ {\bf x} : x_1 \in [-L_1,L_1], x_2 = 0  \}$), 
we have 
\begin{equation} \label{interior_disc}
\nu(\bold{x}_{\bf i}) \, {\ddot{\bf u}}_{\bf i} (t) = {\bold{\mathcal A}}_h ({\bf u}_{\bf i} (t)) + {\tilde{\bf f}}(t,\bold{x}_{\bf i};\boldsymbol{\theta}), \qquad {\bf x}_{\bf i} \in D \setminus \partial D_1,
\end{equation}
where ${\tilde{\bf f}}(t,\bold{x}_{\bf i};\boldsymbol{\theta})$ is a discretization of the 
singular source term, ${\bf f}(t,\bold{x};\boldsymbol{\theta})$, and ${\bold{\mathcal A}}_h$ is 
a difference operator containing standard operators $D_-$, $D_+$, $D_0$ in 
both spatial directions and $\lambda$, $\mu$, and $h$, see \cite{Nilsson_etal:07}. 
On the boundary, $\partial D_1$, we discretize the boundary condition \eqref{elastic_4th},
\begin{equation} \label{absorbing_disc}
{\dot{\bf u}}_{\bf i} (t) = {\bold{\mathcal B}}_h ({\bf u}_{\bf i} (t)), \qquad {\bf x}_{\bf i} \in \partial D_1,
\end{equation}
where ${\bold{\mathcal B}}_h$ is a difference operator containing $D_-$, $D_+$, $\nu$, $\lambda$, $\mu$, and $h$. 

We further collect the semi-discrete solution, ${\bf u}_{\bf i}(t)$, at all $N_h = N_1 \times N_2$ 
grid points in two $N_h$-vectors, $\tilde{u}_1(t)$ and $\tilde{u}_2(t)$. 
We also collect ${\tilde{\bf f}}(t,\bold{x}_{\bf i};\boldsymbol{\theta})$ 
at all $N_h$ grid points in two $N_h$-vectors, 
$\tilde{f}_1(t;\boldsymbol{\theta})$ and $\tilde{f}_2(t;\boldsymbol{\theta})$. 
We finally combine the semi-discrete formulas, \eqref{interior_disc} and \eqref{absorbing_disc},
and write them in matrix form:
\begin{align*}
I_1 \, {\ddot{\tilde{u}}}_1(t) + I_2 \, {\dot{\tilde{u}}}_1(t) &= A_1 \, {\tilde{u}}_1(t) + A_2   \, {\tilde{u}}_2(t) + C \,  \tilde{f}_1(t;\boldsymbol{\theta}),\\
I_1 \, {\ddot{\tilde{u}}}_2(t) + I_2 \, {\dot{\tilde{u}}}_2(t) &= B_1 \, {\tilde{u}}_1(t) + B_2   \, {\tilde{u}}_2(t) + C \,  \tilde{f}_2(t;\boldsymbol{\theta}).
\end{align*}
Here, all matrices, $A_1, A_2, B_1, B_2, C, I_1, I_2$, are in ${\mathbb R}^{N_h \times N_h}$, and  $I_1 + I_2 = I$.

The full discretization is obtained by discretizing the time domain $[0,T]$ into $N_t$ 
equidistant time levels, $t_m \in [0,T]$, where $0 = t_0 < t_1 < \dotsc <t_{N_t-2} < t_{N_t-1}=T$ 
with a time step $\Delta t = T / (N_t -1)$. We let ${\tilde{u}}_1^m$ and ${\tilde{u}}_2^m$ denote 
the full-discrete vectors, ${\tilde{u}}_1(t_m)$ and ${\tilde{u}}_2(t_m)$, respectively, and use the following central difference formulas:
\begin{equation}\label{full_discrete}
{\ddot{\tilde{u}}}_k^m = \frac{{\tilde{u}}_k^{m+1}- 2 \, {\tilde{u}}_k^m + {\tilde{u}}_k^{m-1}}{{\Delta t}^2}, \qquad {\dot{\tilde{u}}}_k^m = \frac{{\tilde{u}}_k^{m+1}-  {\tilde{u}}_k^{m-1}}{2 \, \Delta t}, \qquad k=1,2.
\end{equation}
We finally arrive at the full-discrete formulas
\begin{subequations}\label{Full_Disc}
\begin{align}
U_{1 \, \text{FD}} & \equiv I_1 \, \frac{{\tilde{u}}_1^{m+1}- 2 \, {\tilde{u}}_1^m + {\tilde{u}}_1^{m-1}}{{\Delta t}^2} + I_2 \, \frac{{\tilde{u}}_1^{m+1}-  {\tilde{u}}_1^{m-1}}{2 \, \Delta t} \nonumber\\
&-A_1 \, {\tilde{u}}_1^m - A_2   \, {\tilde{u}}_2^m - C \,  \tilde{f}_1^m = 0,\\
U_{2 \, \text{FD}} & \equiv I_1 \, \frac{{\tilde{u}}_2^{m+1}- 2 \, {\tilde{u}}_2^m + {\tilde{u}}_2^{m-1}}{{\Delta t}^2} + I_2 \, \frac{{\tilde{u}}_2^{m+1}-  {\tilde{u}}_2^{m-1}}{2 \, \Delta t} \nonumber\\
&- B_1 \, {\tilde{u}}_1^m - B_2   \, {\tilde{u}}_2^m - C \,  \tilde{f}_2^m = 0,
\end{align}
\end{subequations}
where $\tilde{f}_k^m = \tilde{f}_k(t_m;\boldsymbol{\theta})$ with $k=1,2$.
\subsubsection{Computation of the Hessian of cost functional}

In this section, we describe in detail how to compute the Hessian,
$\nabla_{\vec \theta} \nabla_{\vec \theta} {\mathcal L}$,
of the cost functional, ${\mathcal L}$, given in \eqref{cost_functional}. 
We consider the first term in the right-hand side of \eqref{cost_functional} and write
\begin{equation}\label{mathbb_L}
{\mathcal L}_1(\vec{\theta}) = \sum_{m=0}^{N_t-1} {\mathbb L}(t_m, {\tilde{u}}_1^m(\vec \theta),{\tilde{u}}_2^m(\vec \theta)), \qquad {\mathbb L} := \frac{1}{2} \sum_{r=1}^{N_R}    {\bf r}_r(t_m,\vec \theta)^{\top} \, {\vec C}_{\epsilon}^{-1} \, {\bf r}_r(t_m,\vec \theta), 
\end{equation}
where the residual ${\bf r}_r$, given in \eqref{res}, 
is a function of ${\tilde u}_1$ and ${\tilde u}_2$, which are 
in turn functions of $t_m$ and $\vec \theta$. The Hessian of the 
remaining terms in the right-hand side of \eqref{cost_functional}, 
i.e. $-h({\vec \theta}) + C$, is simply $-\nabla_{\vec \theta} \nabla_{\vec \theta} h$. 

To obtain the Hessian of ${\mathcal L}_1$, we first introduce the Lagrangian:
\begin{equation}\label{Lagrangian}
{\hat{\mathcal L}}_1 = \sum_{m=0}^{N_t-1} \Bigl( {\mathbb L}(t_m, {\tilde{u}}_1^m(\vec \theta),{\tilde{u}}_2^m(\vec \theta)) + {\varphi_1^m}^{\top} \, U_{1 \, \text{FD}} + {\varphi_2^m}^{\top} \, U_{2 \, \text{FD}}  \Bigr).
\end{equation}
Since by \eqref{Full_Disc}, 
$U_{1 \, \text{FD}} = {\bf 0}$ and $U_{2 \, \text{FD}}= {\bf 0}$, we may choose the Lagrange multipliers $\varphi_1^m$ and $\varphi_2^m$ freely. Consequently, we have ${\hat{\mathcal L}}_1 = {\mathcal L}_1$, and hence $\nabla_{\vec \theta}{\hat{\mathcal L}}_1 = \nabla_{\vec \theta}{\mathcal L}_1$, and $\nabla_{\vec \theta} \nabla_{\vec \theta} {\hat{\mathcal L}}_1 = \nabla_{\vec \theta} \nabla_{\vec \theta} {\mathcal L}_1$. 
In order to avoid long expressions, we set ${\tilde{u}}_{k \, {\vec \theta}}^m := \nabla_{\vec \theta} {{\tilde{u}}_k^m}$ and ${\tilde{f}}_{k \, {\vec \theta}}^m := \nabla_{\vec \theta} {{\tilde{f}}_k^m}$, with $k=1,2$. Thanks to \eqref{Full_Disc}, we have
\begin{align*}
&\nabla_{\vec \theta} {\hat{\mathcal L}}_1 = \sum_{m=0}^{N_t-1} \Bigl( \nabla_{{\tilde{u}}_1^m} {\mathbb L} \,  {{\tilde{u}}_{1 \,{\vec \theta}}^m} +  \nabla_{{\tilde{u}}_2^m} {\mathbb L} \, {{\tilde{u}}_{2 \,  {\vec \theta}}^m} + {\varphi_1^m}^{\top} \, \bigl( I_1 \, \frac{{\tilde{u}}_{1 \, {\vec \theta}}^{m+1}- 2 \, {\tilde{u}}_{1 \, {\vec \theta}}^m + {\tilde{u}}_{1 \, {\vec \theta}}^{m-1}}{{\Delta t}^2}+\\ 
&I_2 \, \frac{{\tilde{u}}_{1 \, {\vec \theta}}^{m+1}-  {\tilde{u}}_{1 \, {\vec \theta}}^{m-1}}{2 \, \Delta t} 
-A_1 \, {\tilde{u}}_{1 \, {\vec \theta}}^m - A_2   \, {\tilde{u}}_{2 \, {\vec \theta}}^m - C \,  \tilde{f}_{1 \, {\vec \theta}}^m \bigr) + {\varphi_2^m}^{\top} \, \bigl(  I_1 \, \frac{{\tilde{u}}_{2 \, {\vec \theta}}^{m+1}- 2 \, {\tilde{u}}_{2 \, {\vec \theta}}^m + {\tilde{u}}_{2 \, {\vec \theta}}^{m-1}}{{\Delta t}^2} \\
&+ I_2 \, \frac{{\tilde{u}}_{2 \, {\vec \theta}}^{m+1}-  {\tilde{u}}_{2 \, {\vec \theta}}^{m-1}}{2 \, \Delta t} 
- B_1 \, {\tilde{u}}_{1 \, {\vec \theta}}^m - B_2   \, {\tilde{u}}_{2 \, {\vec \theta}}^m - C \,  \tilde{f}_{2 \, {\vec \theta}}^m \bigr) \Bigr).
\end{align*}
Using {\it summation by parts}, it is easy to show that the following relations hold for $k=1,2$,
$$
\sum_{m=0}^{N_t-1} {\varphi_k^m}^{\top} \, I_1 \, \frac{{\tilde{u}}_{k \, {\vec \theta}}^{m+1}- 2 \, {\tilde{u}}_{k \, {\vec \theta}}^m + {\tilde{u}}_{k \, {\vec \theta}}^{m-1}}{{\Delta t}^2} = 
\sum_{m=0}^{N_t-1}  \frac{{{\varphi}_{k}^{m+1}}^{\top}- 2 \, {{\varphi}_{k}^m}^{\top} + {{\varphi}_{k}^{m-1}}^{\top}}{{\Delta t}^2} \, I_1 \, {\tilde{u}}_{k \, {\vec \theta}}^m,
$$
and
$$
\sum_{m=0}^{N_t-1} {\varphi_k^m}^{\top} \, I_2 \, \frac{{\tilde{u}}_{k \, {\vec \theta}}^{m+1}-  {\tilde{u}}_{k \, {\vec \theta}}^{m-1}}{2 \, \Delta t} = - \sum_{m=0}^{N_t-1} \frac{{\varphi_k^{m+1}}^{\top}-  {\varphi_k^{m-1}}^{\top}}{2 \, \Delta t} \, I_2 \, {\tilde{u}}_{k \, {\vec \theta}}^m.
$$
Therefore,
\begin{align*}
&\nabla_{\vec \theta} {\hat{\mathcal L}}_1 = \sum_{m=0}^{N_t-1} \Bigl( \bigl( \nabla_{{\tilde{u}}_1^m} {\mathbb L} + \frac{{{\varphi}_{1}^{m+1}}^{\top}- 2 \, {{\varphi}_{1}^m}^{\top} + {{\varphi}_{1}^{m-1}}^{\top}}{{\Delta t}^2} \, I_1 - \frac{{\varphi_1^{m+1}}^{\top}-  {\varphi_1^{m-1}}^{\top}}{2 \, \Delta t} \, I_2\\
& - {{\varphi}_{1}^m}^{\top} \, A_1 - {{\varphi}_{2}^m}^{\top} \, B_1 \bigr) \, {{\tilde{u}}_{1 \,{\vec \theta}}^m}+
\bigl( \nabla_{{\tilde{u}}_2^m} {\mathbb L}
+ \frac{{{\varphi}_{2}^{m+1}}^{\top}- 2 \, {{\varphi}_{2}^m}^{\top} + {{\varphi}_{2}^{m-1}}^{\top}}{{\Delta t}^2} \, I_1 \\
&- \frac{{\varphi_2^{m+1}}^{\top}-  {\varphi_2^{m-1}}^{\top}}{2 \, \Delta t} \, I_2 - {{\varphi}_{1}^m}^{\top} \, A_2 - {{\varphi}_{2}^m}^{\top} \, B_2  \bigr)\, {{\tilde{u}}_{2 \,{\vec \theta}}^m} 
- {{\varphi}_{1}^m}^{\top}  C  \tilde{f}_{1 \, {\vec \theta}}^m
- {{\varphi}_{2}^m}^{\top}  C   \tilde{f}_{2 \, {\vec \theta}}^m  \Bigr).
\end{align*}
Hence, if we let ${\varphi}_{1}^m$ and ${\varphi}_{2}^m$ be the solutions to the following dual problems,
\begin{subequations}\label{Dual_problems}
\begin{align}
{\Phi}_{1 \, \text{FD}} & \equiv  \frac{{{\varphi}_{1}^{m+1}}^{\top}- 2 \, {{\varphi}_{1}^m}^{\top} + {{\varphi}_{1}^{m-1}}^{\top}}{{\Delta t}^2} \, I_1 - \frac{{\varphi_1^{m+1}}^{\top}-  {\varphi_1^{m-1}}^{\top}}{2 \, \Delta t} \, I_2 \nonumber \\ 
&- {{\varphi}_{1}^m}^{\top} \, A_1 - {{\varphi}_{2}^m}^{\top} \, B_1 +\nabla_{{\tilde{u}}_1^m} {\mathbb L} =0,\\
{\Phi}_{2 \, \text{FD}} & \equiv  \frac{{{\varphi}_{2}^{m+1}}^{\top}- 2 \, {{\varphi}_{2}^m}^{\top} + {{\varphi}_{2}^{m-1}}^{\top}}{{\Delta t}^2} \, I_1 
- \frac{{\varphi_2^{m+1}}^{\top}-  {\varphi_2^{m-1}}^{\top}}{2 \, \Delta t} \, I_2 \nonumber\\
&- {{\varphi}_{1}^m}^{\top} \, A_2 - {{\varphi}_{2}^m}^{\top} \, B_2 + \nabla_{{\tilde{u}}_2^m} {\mathbb L}= 0,
\end{align}
\end{subequations}  
then
$$
\nabla_{\vec \theta} {\hat{\mathcal L}}_1 =- \sum_{m=0}^{N_t-1} \Bigl( {{\varphi}_{1}^m}^{\top} C  \tilde{f}_{1 \, {\vec \theta}}^m
+{{\varphi}_{2}^m}^{\top} C \tilde{f}_{2 \, {\vec \theta}}^m  \Bigr).
$$
In a similar way, we can differentiate twice the Lagrangian with respect 
to $\vec \theta$ and use \eqref{Full_Disc} and \eqref{Dual_problems} and summation by parts formulas to obtain
\begin{equation*}
\nabla_{\vec \theta} \nabla_{\vec \theta} {\hat{\mathcal L}}_1 = H_I + H_{II},
\end{equation*}
where
\begin{align}
H_I &= \sum_{m=0}^{N_t-1} \Bigl( ({{\tilde{u}}_{1 \,{\vec \theta}}^m})^{\top} (\nabla_{{\tilde{u}}_1^m} \nabla_{{\tilde{u}}_1^m}  {\mathbb L}) ({{\tilde{u}}_{1 \,{\vec \theta}}^m}) +
({{\tilde{u}}_{2 \,{\vec \theta}}^m})^{\top} (\nabla_{{\tilde{u}}_2^m} \nabla_{{\tilde{u}}_2^m}  {\mathbb L}) ({{\tilde{u}}_{2 \,{\vec \theta}}^m})
 \Bigr), \label{H1_numerics}\\
 H_{II} &= - \sum_{m=0}^{N_t-1}   \sum_{i=1}^{N_h} \Bigl( ({{\varphi}_{1}^m}^{\top} C)_i \nabla_{\vec \theta} \tilde{f}_{1 \, {\vec \theta}}^m(i,:)  +
({{\varphi}_{2}^m}^{\top} C)_i \nabla_{\vec \theta} \tilde{f}_{2 \, {\vec \theta}}^m(i,:) 
 \Bigr). \label{H2_numerics}
 \end{align}
 
Here, $(y)_i \in {\mathbb R}$ means the $i$-th entry of a vector,
$y \in {\mathbb R}^{1 \times N_h}$, and 
${\tilde{f}}_{\vec \theta}(i,:) \in {\mathbb R}^{1 \times N_{\theta}}$ means 
the $i$-th row of a matrix, ${\tilde{f}}_{\vec \theta} \in {\mathbb R}^{N_h \times N_{\theta}}$. 
Note that, in this case, we will have $\nabla_{\vec \theta} \tilde{f}_{{\vec \theta}}(i,:) \in {\mathbb R}^{N_{\theta} \times N_{\theta}}$.
 
  We finally obtain the Hessian of the cost functional, 
\begin{equation}\label{Final_Hessian}
\nabla_{\vec \theta} \nabla_{\vec \theta} {\mathcal L} = H_I + H_{II} - \nabla_{\vec \theta} \nabla_{\vec \theta} h,
\end{equation}
where $H_I$ and $H_{II}$ are given by \eqref{H1_numerics}-\eqref{H2_numerics}. 
The computation of the second term of the Hessian, $H_{II}$, requires 
solving one dual problem \eqref{Dual_problems} and one full elastic 
wave equation \eqref{Full_Disc} to obtain ${{\tilde{u}}_{k}^m}$ and 
consequently $\nabla_{{\tilde{u}}_k^m} {\mathbb L}$, with $k=1,2$. 
However, calculating the first term, $H_I$, requires the 
quantities ${{\tilde{u}}_{1 \,{\vec \theta}}^m}$ and ${{\tilde{u}}_{2 \,{\vec \theta}}^m}$, 
which satisfy the elastic wave equation with the force term ${\bf f}_{\vec \theta}$. 
Therefore, in total, $N_{\theta}+2$ wave equations must be solved to compute the Hessian \eqref{Final_Hessian}.

\begin{rmk}
We note that in practice, we only compute the first part of the
Hessian $H_I$ (see Theorem \ref{TM_4}) for which we need to solve $N_{\theta}$
primal problems. 
If higher accuracy
and consequently the computation of the second part of the Hessian $H_{II}$
is needed (see Theorem \ref{TM_3}), we also need to solve one primal and one dual problem \eqref{Dual_problems}. 
\end{rmk}

In order to further clarify the calculation of the Hessian, we address the following two issues: 

\medskip
\noindent
{\bf 1. Calculation of $\nabla_{{\tilde{u}}_k^m} {\mathbb L}$ and $\nabla_{{\tilde{u}}_k^m} \nabla_{{\tilde{u}}_k^m}  {\mathbb L}$, with $k=1,2$.} By the definition of ${\mathbb L}$ in \eqref{mathbb_L}, we have
$$
\nabla_{{\tilde{u}}_k^m} {\mathbb L} = \sum_{r=1}^{N_R}    {\bf r}_r(t_m,\vec \theta)^{\top} \, 
{\vec C}_{\epsilon}^{-1} \, \nabla_{{\tilde{u}}_k^m}{\bf r}_r(t_m,\vec \theta), 
$$
Note that by \eqref{forward_model} and \eqref{res}, we have
$$
\nabla_{{\tilde{u}}_1^m}{\bf r}_r = 
\begin{pmatrix}
0  & \dotsc & 0 & -1 & 0 & \dotsc & 0  \\
0 & &  & \dotsc & & & 0
\end{pmatrix}, \qquad 
\nabla_{{\tilde{u}}_2^m}{\bf r}_r = 
\begin{pmatrix}
0 & &  & \dotsc & & & 0\\
0  & \dotsc & 0 & -1 & 0 & \dotsc & 0
\end{pmatrix},
$$
which are $2 \times N_h$ matrices with zero elements except one element being $-1$ 
corresponding to the receiver, $r$. Similarly, we obtain 
$\nabla_{{\tilde{u}}_k^m} \nabla_{{\tilde{u}}_k^m}  {\mathbb L} \in {\mathbb R}^{N_h \times N_h}$ 
with zero elements except for $N_R$ diagonal elements being 
$\hat{c}_{11}$ at the locations corresponding to $N_R$ receivers. 
We note that for computing $\nabla_{{\tilde{u}}_k^m} {\mathbb L}$,
we need ${\tilde{u}}_k^m$, but $\nabla_{{\tilde{u}}_k^m} \nabla_{{\tilde{u}}_k^m}  {\mathbb L}$ 
is independent of ${\tilde{u}}_k^m$.

\medskip
\noindent
{\bf 2. Discretization of the singular source function.} We need to discretize 
the source function, ${\bf f}$, in \eqref{source} so that it is twice continuously differentiable with respect to ${\vec \theta}$ (see \eqref{H2_numerics}). 
In other words, the gradient of the Dirac distribution, $\nabla \delta({\bf x} -{\bf x}_s)$, 
needs to be discretized so that it is twice continuously differentiable in ${\bf x}_s$.
For this purpose, we employ the technique proposed in \cite{Sjogreen_Petersson:14} to 
derive regularized approximations of the Dirac distribution and its gradient, which
result in point-wise convergence of the solution away from the sources. The
derivation of approximations of the Dirac distribution and its gradient is based on the following properties:
$$
\int \phi({\bf x}) \, \delta({\bf x} -{\bf x}_s) \, d{\bf x} = \phi({\bf x}_s), \qquad \int \phi({\bf x}) \, \partial_x \delta({\bf x} -{\bf x}_s) \, d{\bf x} = - \partial_x \phi({\bf x}_s),
$$
which hold for any smooth, compactly supported function, $\phi$. In one dimension with a 
uniform grid, $x_k$, with grid size $h$, the integrals are replaced by a discrete scalar 
product, $(p,q)_{1,h} := h \sum p_i \, q_i$. Fourth-order accurate approximations of 
the Dirac distribution and its gradient are for instance obtained when 
the integral conditions are satisfied with $\phi$ being polynomials of 
degree four. Let $x_k \le x_s < x_{k+1}$ and $\alpha = (x_s - x_k)/h$. Then a fourth-order 
discretization of $\delta(x-x_s)$, which is twice continuously differentiable in 
${\bf x}_s$, is given by \cite{Sjogreen_Petersson:14}:
\begin{align*}
&\delta_{k-2} =  (\alpha / 12 - \alpha^2/24-\alpha^3/12 - 19 \alpha^4/24 + P(\alpha)) / h, \\
&\delta_{k-1} =  (-2 \alpha /3 +2 \alpha^2/3+\alpha^3/6 +4 \alpha^4 -5 P(\alpha))/h, \\
&\delta_{k} = ( 1 -5 \alpha^2/4-97 \alpha^4/12  +10 P(\alpha))/h,\\
&\delta_{k+1} = (2 \alpha / 3 +2 \alpha^2/3-\alpha^3/6 + 49 \alpha^4/6 -10 P(\alpha))/h, \\
&\delta_{k+2} =  (- \alpha / 12 - \alpha^2/24+\alpha^3/12 - 33 \alpha^4/8 +5 P(\alpha))/h, \\
&\delta_{k+3} =  ( 5 \alpha^4/6 - P(\alpha))/h, \\
&\delta_{j} = 0, \quad j \notin \{ k-1,k,k+1,k+2 \},
\end{align*}
where $P(\alpha) =5 \alpha^5 / 3 - 7 \alpha^6/24- 17 \alpha^7/12 + 9 \alpha^8/8 - \alpha^9 / 4$. 
Similarly, a fourth-order discretization of $\delta'(x-x_s)$ is given by
\begin{align*}
&\delta'_{k-2} = (-1/ 12 + \alpha / 12+\alpha^2/4 + 2 \alpha^3 / 3 + R(\alpha)) / h^2, \\
&\delta'_{k-1} =  (2/ 3 - 4 \alpha/3-\alpha^2/2 - 7 \alpha^3 / 2 - 5 R(\alpha))  / h^2, \\
&\delta'_{k} =  ( 5 \, \alpha/2  +22 \, \alpha^3/3 + 10 R(\alpha))  / h^2,\\
&\delta'_{k+1} =  (-2/3- 4 \, \alpha/3 +\alpha^2/2 - 23 \, \alpha^3 / 3 - 10 \, R(\alpha))  / h^2, \\
&\delta'_{k+2} =  ( 1/ 12 +\alpha/12 - \alpha^2 / 4 + 4 \, \alpha^3 + 5 \, R(\alpha))  / h^2, \\
&\delta'_{k+3} =  ( -5 \, \alpha^3/6 - R(\alpha))  / h^2, \\
&\delta'_{j} = 0, \quad j \notin \{ k-1,k,k+1,k+2 \},
\end{align*}
where $R(\alpha) = -25 \alpha^4/12 -3 \alpha^5 / 4 +59 \alpha^6/12- 4 \alpha^7 +  \alpha^8$. A 
two-dimensional approximation can for instance be obtained by Cartesian products of one-dimensional discretizations: 
\begin{eqnarray*}
\delta({\bf x} -{\bf x}_s) \approx \delta(x_1 -x_{1s}) \, \delta(x_2 -x_{2s}), \qquad 
\nabla \delta({\bf x} -{\bf x}_s) \approx \left( \begin{array}{c}
\delta'(x_1 -x_{1s}) \, \delta(x_2 -x_{2s})\\
\delta(x_1 -x_{1s}) \, \delta'(x_2 -x_{2s})
\end{array} \right).
\end{eqnarray*}
This representation of the forcing together with the second-order accurate finite difference 
scheme, presented in Section \ref{fdm}, result in an overall second-order convergence of the solution away from the singularity at ${\bf x}_s$ (see also \cite{Motamed_elastic:13}).

The complete algorithm for computing the Hessian is outlined in Algorithm \ref{ALG1}.
\begin{algorithm}
\caption{Calculate the Hessian of the cost functional given a $\vec\theta^*$}
\label{ALG1}
\begin{algorithmic} 
\medskip
\STATE {\bf Calculate $H_I$:}
\medskip
\STATE {\bf 1.} Discretize the source function ${\bf f}$ and obtain $\tilde{f}_k(t;\boldsymbol{\theta}^*)$ for $k=1,2$.
\STATE {\bf 2.} Calculate $\partial_{\theta_j} \tilde{f}_k(t;\boldsymbol{\theta}^*)$ 
for $j=1, \dotsc, N_{\theta}$ by differentiating $\tilde{f}_k(t;\boldsymbol{\theta}^*)$ in step {\bf 1}.
\STATE {\bf 3.} Solve \eqref{Full_Disc} with forces in step {\bf 2} 
and obtain ${{\tilde{u}}_{k \,{\vec \theta}}^m}$ for $k=1,2$.
\STATE {\bf 4.} Find $\nabla_{{\tilde{u}}_k^m} \nabla_{{\tilde{u}}_k^m}  {\mathbb L}$ for $k=1,2$.
\STATE {\bf 5.} Compute $H_I$ by \eqref{H1_numerics}.

\medskip
\medskip

\STATE {\bf Calculate $H_{II}$:}
\medskip
\STATE {\bf 6.} Solve \eqref{Full_Disc} with forces in step {\bf 1} and obtain ${{\tilde{u}}_k^m}$ for $k=1,2$.
\STATE {\bf 7.} Find $\nabla_{{\tilde{u}}_k^m}  {\mathbb L}$ for $k=1,2$.
\STATE {\bf 8.} Solve \eqref{Dual_problems} and obtain $\varphi_k^m$ for $k=1,2$. 
\STATE {\bf 9.} Calculate $\nabla_{\vec \theta} 
\tilde{f}_{k \, {\vec \theta}}^m$ by differentiating $\partial_{\theta_j} 
\tilde{f}_k(t;\boldsymbol{\theta}^*)$ in step {\bf 2}.
\STATE {\bf 10.} Compute $H_{II}$ by \eqref{H2_numerics}.
\end{algorithmic}
\end{algorithm}

\subsubsection{The scaled Hessian}
In seismic source inversion, the parameters span several magnitudes. For example, the time shift and frequency are 
${\cal{O}}\left(1\right)$ and the source momenta are ${\cal{O}}\left(10^{14}\right)$, which potentially lead to 
a Hessian matrix with a large condition number. To deal with this problem, we carry out a change of variables as follows:
\begin{align}
 \tilde{\vec\theta} = \vec S \vec\theta \quad \text{with} \quad \vec S = \sqrt{diag(\vec H_1)}\,. \nonumber
\end{align}
Consequently, the rescaled Hessian reads:
$$
 \tilde{\vec H} = \vec S^{-{\top}} \vec H_1 \vec S^{-1}, 
$$
and the information gain in the new scaled variables is:
\begin{align}
 D_{KL} =& \int_{{\cal{\tilde{\vec\Theta}}}} 
 \log{\left(\frac{p_{{\cal{\tilde{\vec\Theta}}}}(\tilde{\vec\theta} | \{\vec y_r\})}{p_{{\cal{\tilde{\vec\Theta}}}}
 (\tilde{\vec\theta})}\right)}p_{{\cal{\tilde{\vec\Theta}}}}(\tilde{\vec\theta} | \{\vec y_r\})
 d\tilde{\vec\theta} \nonumber \\
 =& -\frac{1}{2}\log((2\pi)^{N_{\theta}} |\tilde{\vec H}(\hat{\vec\theta})^{-1}|) - \frac{N_{\theta}}{2} 
 + \tilde{h}(\hat{\vec\theta}) + \frac{ \tilde{\vec H}(\hat{\vec\theta})^{-1}: \nabla_{\tilde{\vec\theta}}
 \nabla_{\tilde{\vec\theta}} \tilde{h}(\hat{\vec\theta})}{2} 
 + \mathcal O_P(\vec S( \hat{\vec\theta} - \vec\theta^*) ),
\end{align}
%
where $p_{{\cal{\tilde{\vec\Theta}}}}(\tilde{\vec\theta} | \{\vec y_r\}) 
= p_{{\cal{\vec\Theta}}}(\vec S^{-1}\tilde{\vec\theta} | \{\vec y_r\})|\vec S|^{-1} $, 
$p_{{\cal{\tilde{\vec\Theta}}}}(\tilde{\vec\theta} ) 
= p_{{\cal{\vec\Theta}}}(\vec S^{-1}\tilde{\vec\theta} ) |\vec S|^{-1}$, and
$\tilde{h}(\vec\theta) = \log{p_{{\cal{\tilde{\vec\Theta}}}}(\vec\theta)}$. We approximate $\hat{\vec\theta}$ 
by $\vec\theta^*$, such that Theorem \ref{TM_4} can be applied to compute the expected information gain.

\subsubsection{Numerical integration}

In this section, we briefly review two approaches for numerical integration, namely the 
deterministic sparse quadrature and Monte Carlo random sampling. Note 
that we use the sparse quadrature under the assumption that the corresponding 
integrand has a certain level of regularity. We should resort to sampling based 
numerical integration techniques, e.g., Monte Carlo sampling, when this assumption is not valid. 
Due to the singular source term and its twice continuously differentiable discretization, the 
solution of our problem does not possess high regularity with respect to the source location parameters. 

For details of sparse quadratures, see 
\cite{Back:2011a, Barthelmann:2000, Back:2011b, Nobile:2008, Smolyak:1963}.
Also, see \cite{Motamed_elastic:13, Motamed_etal:12} for a detailed regularity analysis 
and convergence study of the sparse quadrature for the stochastic wave equation. 
We can use the interpolating polynomial--based numerical integration and write 
\begin{align}
I = \int_{\vec\Theta} \hat{D}_{KL}(\vec\theta^*) \, p(\vec \theta^*) \, d\vec \theta^*  
= \sum_{i=1}^{\eta} w_i \, \hat{D}_{KL}(\vec\theta^*_i) + {\varepsilon}_{\eta}\, , \label{eq:intpoly} 
\end{align}
where $\{ \vec\theta^*_i \}_{i=1}^{\eta}$ and $\{ w_i \}_{i=1}^{\eta}$ are $\eta$ 
quadrature points and weights, respectively, $\hat{D}_{KL}$ is the approximated 
K--L divergence given by \eqref{DKL_hat}, and $\varepsilon_{\eta}$ is the 
interpolation error. Different types of quadrature rules are available, including Gauss and Clenshaw-Curtis rules. In Gauss quadrature, quadrature points and weights correspond to the multivariate $p$-orthogonal polynomials. For instance, Legendre and Hermite polynomials are used for uniform and Gaussian priors, respectively. In Clenshaw-Curtis quadrature, Chebyshev polynomials are employed to obtain quadrature points and weights. 


In a standard quadrature rule on full tensor product grids, the total number of quadrature 
points, $\eta$, grows exponentially with $N_{\theta}$. Full tensor product approximations 
can therefore be used only when the number of parameters is small (in practice when 
$N_{\theta} \le 3$). In order to suppress the curse of dimensionality, sparse quadrature rules are employed. 
The main strategy used in sparse quadrature is to leave out the fine levels of 
a hierarchical interpolation. A full tensor product rule is recovered if all the 
hierarchical bases are used. Sparse approximations are both accurate and efficient, 
particularly when the integrand, $\hat{D}_{KL}(\vec\theta^*)$, is highly regular with respect to $\vec\theta^*$. 
In general, the following estimates hold for full tensor quadrature rules:
$$
\hat{D}_{KL} \in H^s({\mathbb R}^{N_{\theta}}) \qquad \Rightarrow \qquad \varepsilon_{\eta} = {\mathcal O}(\eta^{-s / N_{\theta}}),
$$
and for sparse quadrature rules:
$$
\hat{D}_{KL} \in H_{\text{mix}}^s({\mathbb R}^{N_{\theta}}) \qquad \Rightarrow \qquad  \varepsilon_{\eta} = {\mathcal O}(\eta^{-s}(\log \eta)^{s \, ({N_{\vec\theta}}-1)}).
$$
Here, $H^s$ is the space of multi-variate functions with square integrable $s>0$ weak 
derivatives with respect to each variable, and $H_{\text{mix}}^s$ is the space of 
multi-variate functions with square integrable $s>0$ {\it mixed} weak derivatives. 
See \cite{NovRit:1997,Motamed_etal:12,Motamed_elastic:13} for more details.  

On the other hand, if we use Monte Carlo random sampling, the expected information gain can be written as
\begin{align}
I = \int_{\vec\Theta} \hat{D}_{KL}(\vec\theta^*) \, p(\vec \theta^*) \, d\vec \theta^*  
= \frac{1}{M}\sum_{j=1}^{M}  \hat{D}_{KL}(\vec\theta^*_j) + {\varepsilon}_M\, , \label{eq:mcs} 
\end{align}
where $\vec\theta^*_j$ is the $j^{th}$ random sample drawn from distribution $p(\vec\theta^*)$, and ${\varepsilon}_M = {\mathcal O_P}
\left(M^{-1/2} \right)$. 

We note that recent advances in Monte Carlo type methods, such as multilevel Monte Carlo \cite{Giles:2008} 
and multi-index Monte Carlo \cite{HajiAli:2014}, can 
be used to accelerate the computation of expected information gain, when the dimension is high and/or the 
integrand function lacks high regularity with respect to the parameters. 

\begin{rmk}
In addition to the quadrature error, $\varepsilon_{\eta}$ or
$\varepsilon_M$, we also need to consider the discretization error in the
computation of Hessian, which is proportional to $h^q$, where $q$ depends
on the order of accuracy of the finite difference scheme and the
regularity of the wave solution. In practice, we take the spatial
grid-length $h$ small enough so that the discretization error does not dominate the quadrature error.
\end{rmk}

\section{Numerical Examples}
\label{sec_numerics}

In this section, we present a few numerical examples to demonstrate the efficiency and
applicability of the numerical method for the fast estimation of the expected information gain described above. 

\subsection{Model problem}
\label{model_numerics}

We consider a layered isotropic elastic material in a two-dimensional space and 
model a simplified earthquake, which is similar to the layer over 
half space problem called LOH.1 \cite{LOH2}. The top layer, $D_{I}$,
extends over $-1000 \le x_2 \le 0$, and the half space, $D_{II}$, is given by $x_2 \le -1000$. 
We truncate the domain and consider the box $D = [-10000,10000] \times [-15000,0]$. 
We impose the stress-free boundary condition on the free surface, $x_2 =0$, and the first-order Clayton-Engquist non-reflecting boundary conditions \cite{Clayton_Engquist:1977} at the artificial boundaries.
The material density and velocities are given by  
\begin{eqnarray*}
\nu({\bf x})=  \left\{ \begin{array}{l l}
2600 &  {\bf x} \in D_I\\
2700 & {\bf x} \in D_{II}
\end{array} \right.
\quad
c_p({\bf x})=  \left\{ \begin{array}{l l}
4000 &  {\bf x} \in D_I\\
6000 & {\bf x} \in D_{II}
\end{array} \right.
\quad
c_s({\bf x})=  \left\{ \begin{array}{l l}
2000 &  {\bf x} \in D_I\\
3464 & {\bf x} \in D_{II}
\end{array} \right.
\end{eqnarray*}
A point moment tensor forcing is applied with a Gaussian time function,
$$
S(t;t_s,\omega_s) = \frac{\omega_s}{\sqrt{2 \, \pi}} \, e^{- \omega_0^2 \, (t - t_s)^2 /2},
$$
which is parametrized by the frequency $\omega_s$ and the center time $t_s$. 
Except for the convergence study in Section \ref{conv_num}, for all experiments, we consider uniform priors for all $N_{\theta}=7$ parameters:
$$
\theta_1\sim {\mathcal U}(-1000,1000), \qquad \theta_2 \sim {\mathcal U}(-3000,-1000), \qquad \theta_3 \sim {\mathcal U}(0.5,1.5), 
$$
$$
\theta_4 \sim {\mathcal U}(3,5), \qquad \theta_5, \theta_6, \theta_7 \sim {\mathcal U}(10^{13}, 10^{15}).
$$
Here, the vector of parameters reads $\boldsymbol{\theta} = (x_{1s}, x_{2s}, t_s, \omega_s, m_{x_1 x_1}, m_{x_1 x_2}, m_{x_2 x_2})^{\top}$.

\begin{figure}[!h]
\centering
\psfrag{x1}{\footnotesize{$x_1$}}
\psfrag{x2}{\footnotesize{$x_2$}}
\psfrag{B1}{\footnotesize{\text{stress-free BC}}}
\psfrag{B2}[c][c][1][90]{\footnotesize{\text{non-reflecting BC}}}
\psfrag{B3}[c][c][1][270]{\footnotesize{\text{non-reflecting BC}}}
\psfrag{B4}[c]{\footnotesize{\text{non-reflecting BC}}}
\includegraphics[width=0.45\textwidth]{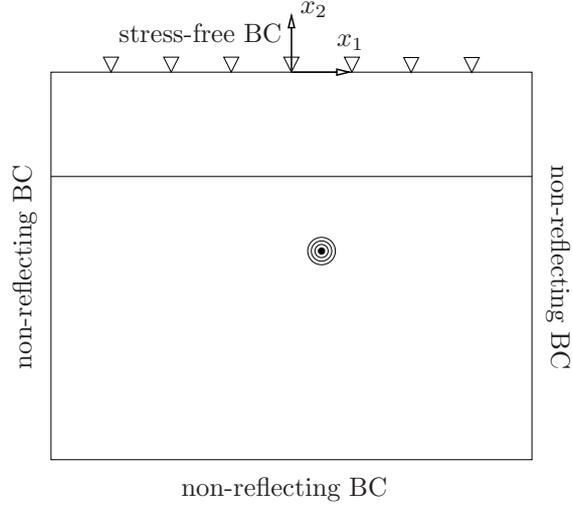}
\caption{The two-layered spatial domain $D = [-10000,10000] \times [-15000,0]$ with stress-free and non-reflecting boundary conditions. An array of $N_R$ receivers are located on the ground surface in equidistant recording points.}
\label{numerics_domain}
\end{figure}

\begin{figure}[!h]
\centering
\includegraphics[width=0.7\textwidth]{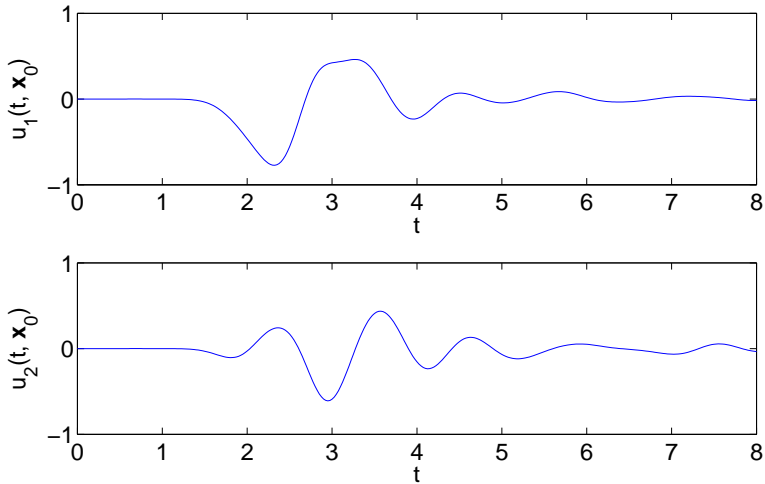}
\caption{The ground motions versus time at the receiver station,
${\bf x}_0=(5000,0)$. The motions are due to solving the elastic wave equation with 
the prior source parameters, ${\mathbb E}[ \boldsymbol{\theta}] = (0,-2000,1,4,10^{14},10^{14},10^{14})^{\top}$.}
\label{example}
\end{figure}

The array of receivers is placed on the ground surface. See Figure \ref{numerics_domain}. The observation vector contains all displacements measured at the receivers. We assume that the measurement errors for the horizontal and vertical displacements at each receiver  are independent Gaussian random variables,
$\vec \epsilon_r \sim \mathcal{N}(\vec 0, \vec {\bf C}_{\epsilon})$,
with $\vec {\bf C}_{\epsilon} =\begin{pmatrix}
0.0001 & 0  \\
0 & 0.0001
\end{pmatrix}$ for all receivers, $r=1, \dotsc, N_R$. We employ the second-order finite 
difference approximation proposed in \cite{Nilsson_etal:07} with a fixed spatial grid-length,
$h=200$, and a time step, $\Delta t = 0.025$. The elastic wave equation is integrated to the time,
$T=8$. We note that $h=200$ is the largest spatial grid-length for which the
discretization error, which is proportional to ${\mathcal O}(h^2)$, does not dominate the quadrature
error in the integration of the information gain \eqref{eq:mcs} with respect to the prior distribution. We record the wave solutions at $N_t= 1+ T / \Delta t = 321$ discrete time levels. 
For example, Figure \ref{example} shows the ground motions at the receiver station,
${\bf x}_0=(5000,0)$, as a function of time, due to solving the elastic wave equation 
with the prior source parameters, ${\mathbb E}[ \boldsymbol{\theta}] = (0,-2000,1,4,10^{14},10^{14},10^{14})^{\top}$. 

\subsection{The scaled Hessian}

The sizes of the source parameters span many orders of magnitude. In SI units, we have
$$
\theta_1, \theta_2 = {\mathcal O}(10^4)\, \text{m}, \quad \theta_3 =  
{\mathcal O}(1)\, \text{s}, \quad \theta_4 =  {\mathcal O}(10)\, \text{1/s}, 
\quad \theta_5, 
\theta_6, \theta_7 = {\mathcal O}(10^{13}) - {\mathcal O}(10^{15})\, \text{Ns}.
$$ 
Consequently, the condition number of the Hessian is very large. The Hessian therefore is scaled, as described in Section 3.3.3. For more 
clarification, we compute the Hessian matrix, $H_1$, in \eqref{H1_numerics} for the 
expected value of the prior source parameters,
${\mathbb E}[ \boldsymbol{\theta}]= (0,-2000,1,4,10^{14},10^{14},10^{14})^{\top}$. The 
condition number of the unscaled Hessian computed by {\sc{Matlab}} is {\it cond}
($\vec H_1$) $=  3.88 \times 10^{30}$. Now, if we use the scaling matrix, 
$\vec S \in {\mathbb R}^{7 \times 7}$, with diagonal elements, $S_{ii} = \sqrt{ H_{1_{ii}}}$,
and zero off-diagonal elements, then the condition number of the scaled Hessian,
$\hat{\vec H_1} = \vec S^{-\top} \vec H_1 \vec S^{-1}$, reduces significantly to {\it cond}($\hat{\vec H_1}$) = 12.16.

\subsection{Convergence of sparse quadrature}
\label{conv_num}

\begin{figure}[!h]
\centering
\includegraphics[width=0.7\textwidth]{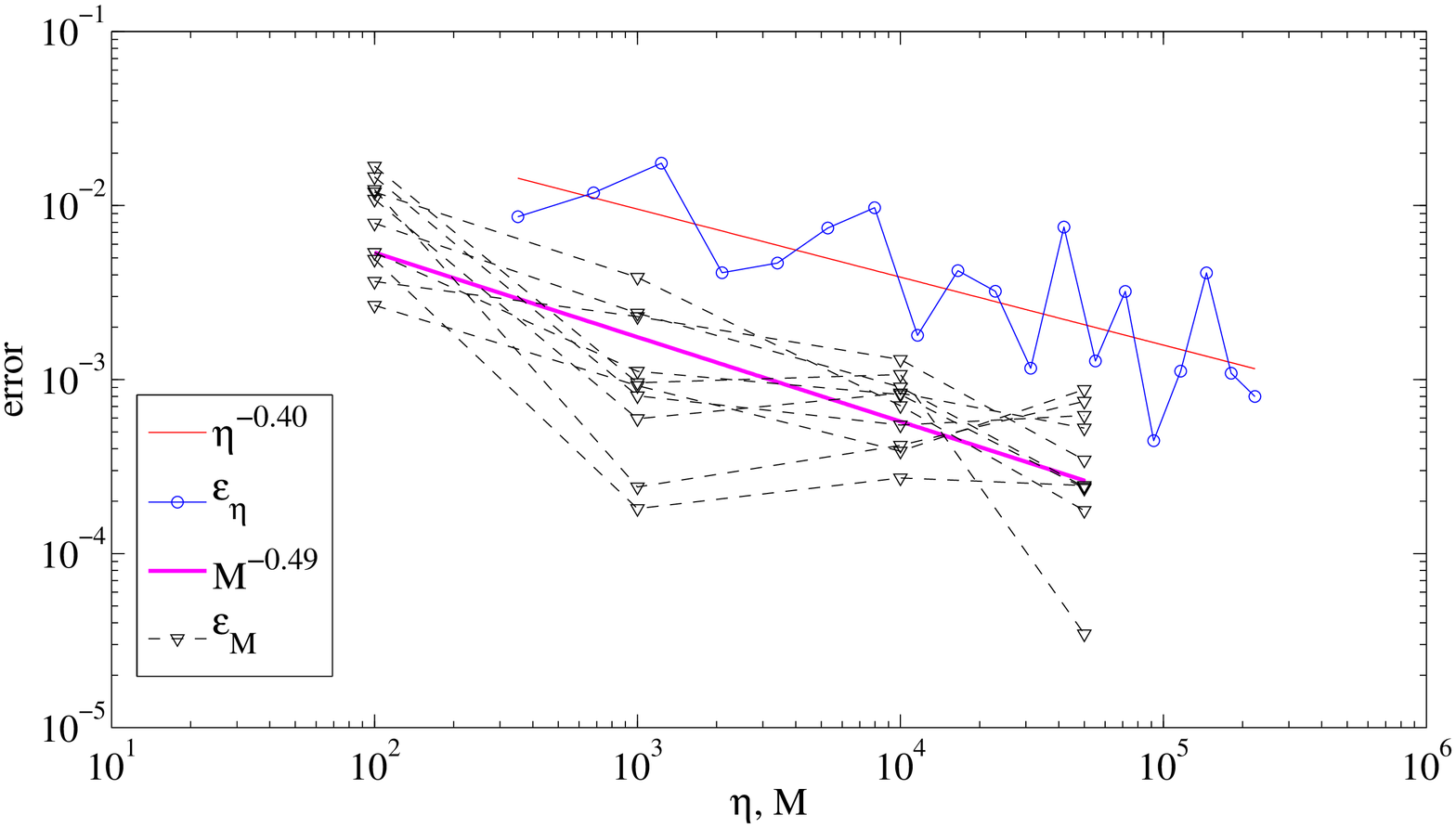}
\caption{The relative errors in sparse quadrature and Monte Carlo versus the number of quadrature points and samples. The rate of convergence, obtained by linear regression through the data points, is $0.40$ for sparse quadrature and $0.49$ for Monte Carlo sampling.}
\label{conv_study}
\end{figure}

In this section, we numerically study the convergence of the two quadrature techniques based on sparse grids and Monte Carlo samples. We consider three parameters: 
$$
\theta_2  \sim {\mathcal U}(-3000,-1000), \qquad \theta_4 \sim {\mathcal U}(3,5), \qquad 
\theta_5 \sim {\mathcal U}(10^{13}, 10^{15})
$$
and leave the other four parameters fixed, i.e. $\theta_1 = -1000, \theta_3 = 1$, and $\theta_6 = \theta_7 = 10^{14}$. 
We use a  Gauss-Legendre sparse grid based on a total degree multi-index set. See \cite{Long:2013} for details on the construction of sparse grids. We consider a sequence of twenty sparse grids with $\eta_i = 351, 681, \dotsc, 271857$ quadrature points. 
The sparse grids correspond to three directions and are obtained by total degree index sets. We then compute the relative error,
$$
\varepsilon_{\eta_i} = \frac{| I_{\eta_{i+1}} - I_{\eta_{i}} |}{| I_{\eta_{i+1}} |}, \qquad i=1, \dotsc, 19. 
$$ 
We also consider a sequence of $M_i=10^2, 10^3, 10^4, 5 \times 10^4, 10^5$ random samples and carry on ten realizations of each $M_i$. In a similar way as above, we compute the relative error $\varepsilon_{M_i}$ in the Monte Carlo sampling technique. Figure \ref{conv_study} shows the relative errors in sparse quadrature $\varepsilon_{\eta}$ and in Monte Carlo sampling technique $\varepsilon_{M}$, versus 
the number of quadrature points $\eta$ and the number of samples $M$. A simple linear regression through the data points shows that the rate of convergence of sparse quadrature is $0.40$, while the rate of convergence of Monte Carlo is $0.49$. The slow rate of convergence in sparse quadrature is a result of low regularity of $I$ with respect to $\boldsymbol \theta$. Specifically, the solution of our problem does not have high regularity with respect to the parameters of source location, due to the singular source term. However, the regularity needed to satisfy Assumptions 1 and 2 can still be provided by the twice continuously differentiable discretization of the delta function in the source term.

\subsection{Comparison of Laplace method and nested Monte Carlo Sampling}
\label{comparison_la_dlmc}

\begin{figure}[!h]
\centering
\includegraphics[scale=0.6]{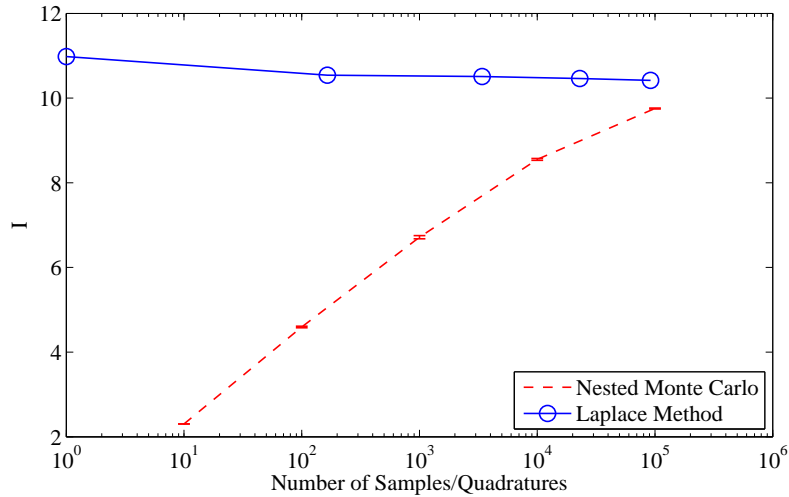}
\caption{Comparison of the convergence performances of Laplace method and nested Monte Carlo sampling. }
\label{comp}
\end{figure}

We numerically verify the concentration of measure by comparing the results of Laplace method (sparse quadrature) and direct nested Monte Carlo sampling. We assume three parameters are known with values identical to those in the previous subsection. We collect data for a period of $T=1.25$ at two receivers located at $x_1 = -9000$ and $x_1 = 1000$. The values of expected information gains computed by both methods with respect to the number of samples/quadratures are shown in Figure \ref{comp}. Note that the Laplace method 
converges much faster than the nested Monte Carlo and the difference between the final results is less than $4\%$.  The strong bias in the nested Monte Carlo is due to the fact that we reused the samples in the inner and outer loops. 

\subsection{Experimental setups}

We carry out three sets of experiments for the model problem in Section \ref{model_numerics}:
\begin{itemize}
\item {\bf Scenario I}: The number of receivers and the distance between the receivers vary, but the interval on which the receivers are distributed evenly and symmetrically around $x_1 = 0$ is fixed, [-8000,8000]. In particular, we consider the following settings:
\begin{center}
\begin{tabular}{c|c c c c c c}
$N_R$     & 3 & 5 & 9 & 17 & 41 & 81 \\
\hline
$d_R$      & 8000 & 4000 & 2000 & 1000 & 400 & 200 \\
\end{tabular}
\, ,
\end{center}
giving a total of six experiments. $d_R$ is the distance between two consecutive receivers.

\item {\bf Scenario II}: The number of receivers varies, $N_R = 1, 3, 5, \dotsc, 19$, and 
the distance between the receivers is fixed, $d_R = 1000$. We distribute the receivers 
evenly and symmetrically around $x_1 = 0$. This gives a total of 10 experiments.

\item {\bf Scenario III}: The number of receivers is fixed, $N_R = 5$, and the distance 
between the receivers varies, $d_R = 200, 400, 600, \dotsc, 4000$. We distribute the receivers
evenly and symmetrically around $x_1 = 0$. This gives a total of 20 experiments.

%
\end{itemize}

\begin{figure}[!h]
\centering
\includegraphics[width=0.7\textwidth]{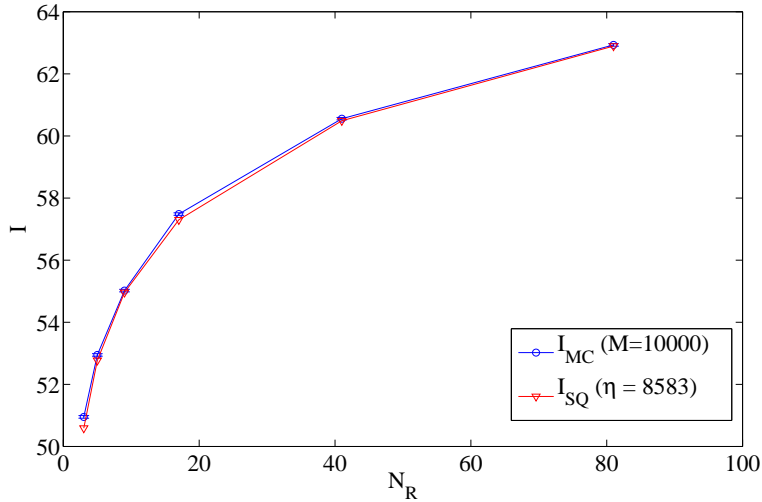}
\caption{The expected information gain, computed both by 
Monte Carlo sampling with $M=10^4$ samples (together with $68.27\%$ confidence interval) 
and by sparse quadrature with $\eta=8583$ quadrature points, 
for six experiments in scenario I. The confidence intervals are indeed very small, less than $1\%$.}
\label{scen3}
\end{figure}

Figure \ref{scen3} shows the expected information gain, computed 
both by Monte Carlo sampling with $M=10^4$ samples and by sparse 
quadrature with $\eta=8583$ quadrature points, for six experiments 
in scenario I. The expected information gain increases sharply until the number of seismograms reaches $20$. 
The extra gains of information is marginal when the number of seismograms is more than $20$,

\begin{figure}[!h]
\centering
\includegraphics[width=0.7\textwidth]{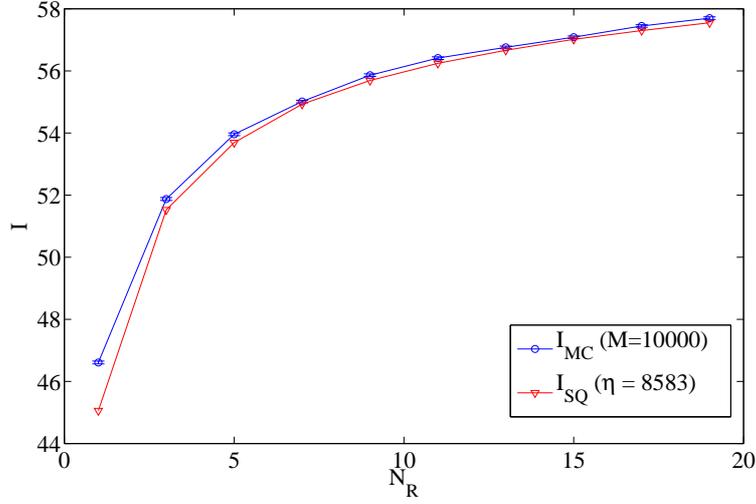}
\caption{The expected information gain, computed both by Monte Carlo 
sampling with $M=10^4$ samples (together with $68.27\%$ confidence interval) and by sparse 
quadrature with $\eta=8583$ quadrature points, for 10 experiments in scenario II.}
\label{scen2}
\end{figure}

Figure \ref{scen2} shows the expected information gain, computed both by Monte Carlo 
sampling with $M=10^4$ samples and by sparse quadrature with $\eta=8583$ quadrature 
points, for 10 experiments in scenario II. It shows that as we increase the number 
of seismograms, the information gain increases marginally.

\begin{figure}[!h]
\centering
\includegraphics[scale = 0.45]{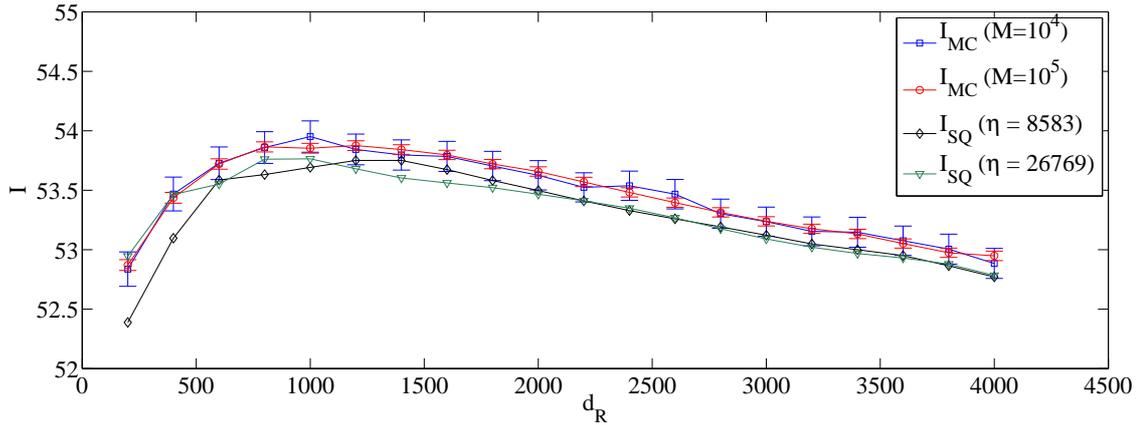}
\caption{The expected information gain, computed both by Monte Carlo 
sampling (together with $99.7\%$ confidence interval) and by sparse quadrature, for 20 experiments in scenario III.}
\label{scen1}
\end{figure}

Figure \ref{scen1} shows the expected information gain, computed both by Monte Carlo 
sampling with $M= 10^4$ and $M=10^5$ samples and by sparse quadrature with 
$\eta=8583$ and $\eta= 26769$ quadrature points, for 20 experiments in scenario III. 
It shows that the experiment with $d_R = 1000$ gives the maximum information. Both lumping and sparsifying the seismograms give suboptimal designs.

\begin{figure}[!h]
\centering
\includegraphics[width=0.7\textwidth,height=6.5cm]{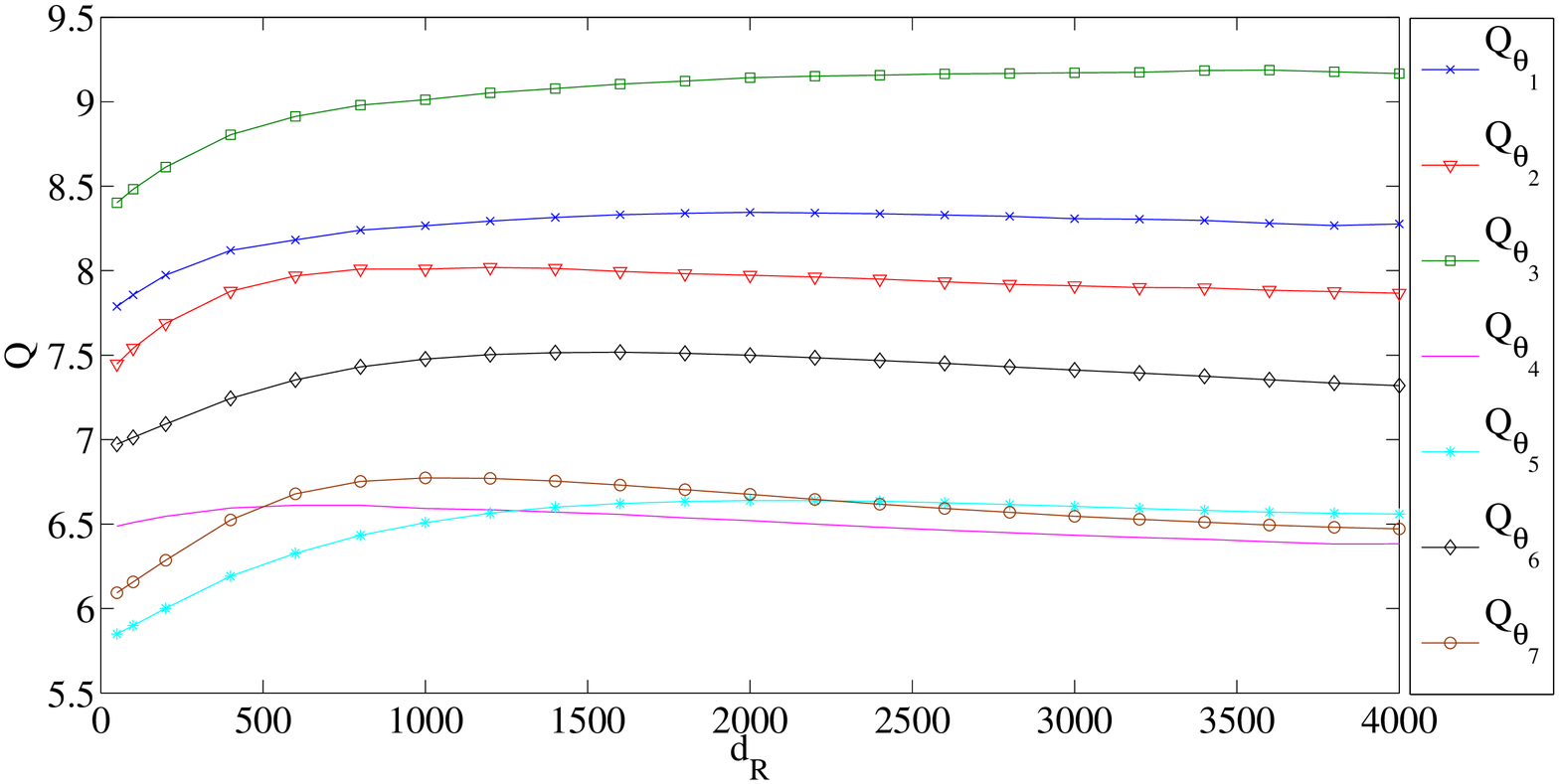}
\caption{The expected information gain corresponding to each parameter 
separately, computed by Monte Carlo sampling with $M=10^5$ samples for 20 experiments in scenario III.}
\label{scen1_QoI}
\end{figure}

Figure \ref{scen1_QoI} shows seven quantities of interest, 
${\mathcal Q}_{\theta_i}$ with $i=1, \dotsc, 7$, which represent the 
information gains of each parameter separately, computed by Monte Carlo 
sampling with $M=10^5$ samples for 20 experiments in scenario III. 
The experiment with approximately $d_R = 1000$ gives the maximum information for $\theta_2$, $\theta_6$ and $\theta_7$.
The experiment with approximately $d_R = 500$ gives the maximum information for $\theta_4$.
The experiment with approximately $d_R = 2000$ gives the maximum information for $\theta_5$.
However, sparsifying the seismograms does not induce a drop of information gain in $\theta_1$ and $\theta_3$. 

Note that we simply sweep over the design spaces to search for the optimal designs because all the scenarios we considered are one-dimensional with respect to the experimental setup, $\xi$. In the cases that more freedom is allowed in a higher dimensional design space, more advanced optimisation algorithm should be implemented.

\section{Conclusion}

We have developed a fast method of optimal experimental design for statistical seismic source 
inversion in a Bayesian setting. 
This method can be generally applied to the design of non-repeatable experiments with a time-dependent model, as long as the 
assumptions in Section 3 are fulfilled. Taking into account that the Hessian of the cost functional
is proportional to the product of the number of points in the time series of the measurements and the number of receivers, we use Laplace approximation to derive an analytical form of the 
information gain, which is a function of the determinant of the aforementioned Hessian matrix. 
The expected information gain eventually reduces to a
marginalization of the information gain over all possible values of the unknown source parameters. 
The asymptotic error terms have been derived. We have applied the new technique to the optimal design of the number and location of seismic receivers on the ground for a simplified two-dimensional earthquake.

\section*{Acknowledgements}
The authors are thankful for support from the Academic Excellency Alliance UT
Austin-KAUST project--Uncertainty quantification for predictive modeling of the dissolution of 
porous and fractured media. Quan Long and Raul Tempone are members of the KAUST SRI Center for Uncertainty 
Quantification in Computational Scinece and Engineering.

\section*{Appendix A.}

In this appendix, we show that 
$$
 \sum_{r=1}^{N_R} \, \sum_{m=0}^{N_t-1} \,  {\vec{\epsilon}_r}^{\top} \,  {\vec C}_{\epsilon}^{-\top}  \, \nabla_{\vec\theta} \vec g_r(t_m, {\vec{\theta}}^*) = {\mathcal O}_P({N}^{1/2}).
$$
For simplicity and to avoid tensor notations, we consider only the one-dimensional case, i.e.,
$d=1$. The case when $d \ge 2$ follows in a similar way and with no difficulty. 

We first note that when $d=1$, ${\vec g}_r(t_m,{\vec\theta}) = u(t_m,x_r ; {\vec\theta})$, $\vec{\epsilon}_r= \epsilon_r$, and ${\vec C}_{\epsilon} = c_{\epsilon}$ are scalar quantities. 
Now, let $u(t_m,x; {\vec\theta})$ be discretized on a spatial grid with $N_h$ grid points. We collect $u$ on all grid points in a $N_h$-vector denoted by $\tilde{u}(t_m,{\vec\theta})$. By the chain rule, we have
$$
\nabla_{\vec\theta} u(t_m,x_r ; {\vec\theta}^*) =\nabla_{\tilde{u}}  u(t_m,x_r ; {\vec\theta}^*)  \, \nabla_{\vec\theta} \tilde{u}(t_m, {\vec\theta}^*).
$$
Moreover, since $\partial_{u(t_m,x ; {\vec\theta})}  u(t_m,x_r ; {\vec\theta}) = \delta(x - x_r)$, then
$$
\nabla_{\tilde{u}}  {u}(t_m,x_r ; {\vec\theta}) = [0,\dotsc, 0, 1 , 0, \dotsc, 0] =: {\mathbb I}_r,
$$   
where ${\mathbb I}_r$ denotes a $1 \times N_h$ vector
whose elements are zero, except a 1 at the position of the $r$-th recorder 
in the grid, denoted by $j(r)$. Therefore,
$$
\sum_{r=1}^{N_R} \, \sum_{m=0}^{N_t-1} \,  {\vec{\epsilon}_r}^{\top} \,  {\vec C}_{\epsilon}^{-\top}  \, \nabla_{\vec\theta} \vec g_r(t_m, {\vec{\theta}}^*) =
\sum_{r=1}^{N_R}  \frac{\epsilon_r}{c_{\epsilon}}  \, {\mathbb I}_r \, \sum_{m=0}^{N_t-1}\,   \nabla_{\vec\theta} \tilde{u}(t_m, {\vec\theta}^*). 
$$
Note that $\nabla_{\vec\theta} \tilde{u}$ is a $N_h \times N_{\theta}$ matrix, and 
$\sum_{r=1}^{N_R}  (\epsilon_r / c_{\epsilon}) \, {\mathbb I}_r$ is a $1 \times N_h$ 
vector, whose elements are zero, except at $N_R$ positions $\{ j(r) \}_{r=1}^{N_R}$, 
corresponding to the ${N_R}$ recording points. Then, the right-hand side in the above formula 
is a $1 \times N_{\theta}$ vector, whose $i$-th element reads $\sum_{r=1}^{N_R}  (\epsilon_r / c_{\epsilon}) \, \sum_{m=0}^{N_t-1}\, \partial_{{\vec\theta}_i} \tilde{u}_{j(r)}(t_m, {\vec\theta}^*)$, with $i=1,\dotsc,N_{\theta}$. 
The desired estimate follows, noting that $\sum_{m=0}^{N_t-1}\, |\partial_{{\vec\theta}_i} \tilde{u}_{j(r)}|$ is bounded from below and above away from zero, thanks to assumptions A1 and A2.

\section*{Appendix B.}

In this appendix, we show that 
$$
\sum_{r=1}^{N_R} \, \sum_{m=0}^{N_t-1} \,  \nabla_{\vec\theta} \nabla_{\vec\theta} 
\vec g_r(t_m, {\vec{\theta}}^*) \circ {\vec C}_{\epsilon}^{-1} \,  \vec{\epsilon}_r  = {\mathcal O}_P({N}^{1/2}).
$$ 
As in appendix A, we consider only the case when $d=1$. We have
$$
\nabla_{\vec\theta}  \nabla_{\vec\theta} {\vec g}_r = \nabla_{\vec\theta} \Bigl( {\mathbb I}_r \, \nabla_{\vec\theta} \tilde{u} \Bigr) = \nabla_{\vec\theta}  \nabla_{\vec\theta} \tilde{u} \circ {\mathbb I}_r^{\top},
$$
where $\nabla_{\vec\theta}  \nabla_{\vec\theta} \tilde{u}$ is a 
$N_{\theta} \times N_{\theta} \times N_h$ tensor. Note that for 
two real vectors, ${\bf a},{\bf b} \in {\mathbb R}^{p}$, we use the notation $\nabla_{\vec\theta} \nabla_{\vec\theta} {\bf a} \circ {\bf b} = \sum_{i=1}^{p}b_i \, \nabla_{\vec\theta} \nabla_{\vec\theta} a_i$. Therefore, we have 
$$
\sum_{r=1}^{N_R} \, \sum_{m=0}^{N_t-1} \,  \nabla_{\vec\theta} \nabla_{\vec\theta} \vec g_r(t_m, {\vec{\theta}}^*) \circ {\vec C}_{\epsilon}^{-1} \,  \vec{\epsilon}_r  =
\sum_{m=0}^{N_t-1} \, \nabla_{\vec\theta}  \nabla_{\vec\theta} \tilde{u}(t_m, {\vec\theta}^*) \circ \sum_{r=1}^{N_R} {\mathbb I}_r^{\top} \, \frac{{\epsilon}_r}{c_{\epsilon}}. 
$$
Similar to Appendix A, we can show that the right hand side in the above formula is 
a $N_{\theta} \times N_{\theta} $ matrix whose element $(i,k)$ reads 
$\sum_{r=1}^{N_R}  (\epsilon_r / c_{\epsilon}) \, \sum_{m=0}^{N_t-1}\, 
\partial_{{\vec\theta}_i {\vec\theta}_k}^2 \tilde{u}_{j(r)}(t_m, {\vec\theta}^*)$. 
The desired estimate follows, noting that $\sum_{m=0}^{N_t-1} 
\partial_{{\vec\theta}_i {\vec\theta}_k}^2 \tilde{u}_{j(r)}$ is bounded 
from below and above away from zero, thanks to assumptions A1 and A2.

\section*{Appendix C.}

In this appendix, we show that 
$$
\sum_{r=1}^{N_R} \, \sum_{m=0}^{N_t-1}  \,  \nabla_{\vec\theta} 
\vec g_r(t_m, {\vec{\theta}}^*)^{\top} \, {\vec C}_{\epsilon}^{-1} \,  
\nabla_{\vec\theta} \vec g_r(t_m, {\vec{\theta}}^*) = {\mathcal O}(N).
$$
Again, we consider only the case when $d=1$. Since $\nabla_{\vec\theta} 
\vec g_r = {\mathbb I}_r \,  \nabla_{\vec\theta} \tilde{u}$, we have
$$
 \sum_{r=1}^{R} \, \sum_{m=0}^{N_t-1}   \nabla_{\vec\theta} 
 \vec g_r(t_m, {\vec\theta}^*)^{\top} \, {\vec C}_{\epsilon}^{-1} \, 
 \nabla_{\vec\theta} \vec g_r(t_m, {\vec\theta}^*) = c_{\epsilon}^{-1} \, 
\sum_{m=0}^{N_t-1} \nabla_{\vec\theta} \tilde{u}(t_m, {\vec\theta}^*)^{\top} 
\, \sum_{r=1}^{N_R}  {\mathbb I}_r^{\top} \, {\mathbb I}_r \,  \nabla_{\vec\theta} \tilde{u}(t_m, {\vec\theta}^*).
$$
We note that $\sum_{r=1}^{N_R}  {\mathbb I}_r^{\top} \, {\mathbb I}_r$ is 
a diagonal $N_h \times N_h$ matrix whose diagonal elements are zeros, 
except elements $\{ (j(r),j(r)) \}_{r=1}^{N_r}$, which are 1's. Therefore, 
the right-hand side in the above formula is a $N_{\theta} \times N_{\theta}$
matrix whose element $(i,k)$ reads $ c_{\epsilon}^{-1} \, 
\sum_{r=1}^{N_R} \, \sum_{m=0}^{N_t-1}\, \partial_{{\vec\theta}_i}  \tilde{u}_{j(r)} 
\, \partial_{{\vec\theta}_k}  \tilde{u}_{j(r)}$. 
The desired estimate follows, noting that $\sum_{m=0}^{N_t-1}\, |\partial_{{\vec\theta}_i}  \tilde{u}_{j(r)} \, \partial_{{\vec\theta}_k}  \tilde{u}_{j(r)}|$ is bounded from below and above away from zero, thanks to assumptions A1 and A2.

\end{document}